\documentclass[a4paper]{aa}
\usepackage{txfonts}
\usepackage{epsfig,amssymb}

\newcommand{\ha}{\ensuremath{\mathrm{H}\alpha}}
\newcommand{\ie}{\emph{i.e.} }
\newcommand{\ebv}{\ensuremath{\mathrm{E}(\mathrm{B}-\mathrm{V})}}

\begin{document}

\title{Low Mass Pre-Main Sequence stars in the Large Magellanic Cloud
- II: HST-WFPC2 observations of two fields in the 30~Doradus region
\thanks{Based on observations with the NASA/ESA \emph{Hubble Space
Telescope}, obtained at the Space Telescope Science Institute, which
is operated by AURA, Inc.,under NASA contract NAS 5-26555.}  }

\author {Martino Romaniello \inst{1}
\and Salvatore Scuderi \inst{2}
\and Nino Panagia \inst{3}$^,$\inst{4}
\and Rosa Maria Salerno \inst{5}
\and Carlo Blanco \inst{5}}

\institute{European Southern Observatory, Karl-Schwarzschild-Strasse 2,
D--85748 Garching bei M\"{u}nchen, Germany; \\{\it mromanie@eso.org}.
\and Osservatorio Astrofisico di Catania, Via S. Sofia 78, 
I--95123 Catania, Italy; {\it scuderi@ct.astro.it}.
\and Space Telescope Science Institute, 3700 San Martin Drive,  
     Baltimore, MD 21218, USA; {\it panagia@stsci.edu}.
\and Affiliated to the Astrophysics Division, Space Science 
     Department of ESA.
\and Dipartimento di Fisica e Astronomia, Via S. Sofia 78, I--95123 Catania.}

\date{Received... Accepted...}

\abstract{As a part of an ongoing effort to characterise the young
stellar populations in the Large Magellanic Cloud, we present
HST-WFPC2 broad and narrow band imaging of two fields with recent star
formation activity in the Tarantula region. A population of objects
with \ha\ and/or Balmer continuum excess was identified. On account of
the intense \ha\ emission (equivalent widths up to several tens of
\AA), its correlation with the Balmer continuum excess and the stars'
location on the HR diagram, we interpret them as low mass
($\sim1-2~M_{\sun}$) Pre-Main Sequence stars.  In this framework, the
data show that coeval high and low mass stars have significantly
different spatial distributions, implying that star formation
processes for different ranges of stellar masses are rather different
and/or require different initial conditions. We find that the overall
slope of the mass function of the young population is somewhat steeper
than the classical Salpeter value and that the star formation density
of this young component is
$0.2-0.4~M_{\sun}\mathrm{yr}^{-1}\mathrm{kpc}^{-2}$, \ie intermediate
between the value for an active spiral disk and that of a starburst
region. The uncertainties associated with the determination of the
slope of the mass function and the star formation density are
thoroughly discussed.   }

\titlerunning{Pre-Main Sequence stars in the LMC}
\maketitle
     
\section{Introduction\label{sec:intro}}
Star-forming processes determine much of the appearance of the visible
Universe: the shape of the Initial Mass Function (IMF) and its
normalisation (the star-formation rate) are, together with stellar
evolution theory, key ingredients in determining the chemical
evolution of a galaxy and its stellar content. Yet, our theoretical
understanding of the processes that lead from diffuse molecular clouds
to stars is still very tentative, as many complex physical phenomena
are simultaneously at play in producing the final results.

From an observational standpoint, most of the effort was traditionally
devoted to nearby Galactic star-forming regions. If, on the one hand,
studying them leads to the the best possible angular resolution, on
the other hand, this is achieved at the expense of probing only a very
limited set of astrophysical conditions (all these clouds have
essentially solar metallicity, e.g. Padget~\cite{padg96}). Also, they
typically cover large areas on the sky that require extensive tiling
to be fully covered with the current detectors (the Taurus star
forming region, for example, covers as much as 12 degrees on the sky).

However, studying the effects of lower metallicity on star formation
is essential to understand the evolution of both our own Galaxy, in
which a large fraction of stars were formed at metallicities below
solar, and of what is observed at high redshifts. As a matter of fact,
the global star formation rate appears to have been much more vigorous
(a factor of 10 or so) at $z\simeq 1.5$ than it is today (Madau et
al~\cite{mad96} and subsequent incarnations of the so-called ``Madau
plot'').  At this epoch the mean metallicity of the interstellar gas
was similar to that of the Large Magellanic Cloud (LMC, e.g. Pei, Fall
\& Hauser~\cite{pei99}). This fact combined with the galaxy's
well-determined distance and small extinction makes the study of star
forming regions in the LMC an important step towards understanding the
general picture of galaxy evolution.

With a distance modulus of $18.57\pm0.05$ (see the discussion in
Romaniello et al~\cite{rom00}), the LMC is the closest galactic
companion to the Milky Way after the Sagittarius dwarf galaxy. At this
distance one arcminute corresponds to roughly 15~pc and, thus, one
pointing with most of the current generation of instruments
comfortably covers almost any individual star forming region in the
LMC (see, e.g., Hodge~\cite{hod88}). In particular, the field of view
of $2.5\arcmin\times2.5\arcmin$ of the camera we have used, the WFPC2
on board the HST, corresponds to $37\times37~\mathrm{pc}^2$ and leads
to the detection of several thousands of stars per pointing. The LMC
is especially suited for stellar populations studies for two
additional reasons. First, its depth along the line of sight is
negligible, at least in the central parts we consider (van der Marel
\& Cioni~\cite{mar01}), and all the stars can be effectively
considered at the same distance, thus minimising a spurious scatter in
the Colour-Magnitude Diagrams (CMDs). Second, the extinction in its
direction due to dust in our Galaxy is low, about $\ebv\simeq 0.05$
(Bessell~\cite{bes91}, Schwering~\cite{sch91}), and hence, our view is
not severely obstructed.

The advent of the Hubble Space Telescope (HST) has for the first time
made it possible to observationally tackle the open questions about
star formation in outer galaxies and, in particular, in the LMC, down
to a solar mass or even lower. Some of the evidence from ground and
HST-based studies shows that there may be significant differences
between star formation processes in the LMC and in the Galaxy.
Immediately after the first HST refurbishment mission, the
observations of the double cluster NGC~1850 in the LMC made with Wide
Field Camera 2 (WFPC2) have provided the first detection of a
population of Pre-Main-Sequence (PMS) stars (Gilmozzi et al 1994) in
an extragalactic star forming region.  The evidence was purely
statistical, in the sense that the existence of PMS stars was deduced
by the presence of many stars that were lying above the main sequence
and whose number (almost 400) could not be accounted for by any
evolved population.  Subsequently, Romaniello (1998), Panagia et al.
(2000) and Romaniello et al (2002), by comparison of the magnitudes
obtained with an $H\alpha$ narrow band filter (F656N) with those taken
with a broad band red filter(F675W), measured the equivalent width of
the $H\alpha$ emission for several hundred of candidates PMS stars in
the field around SN 1987A, thus confirming their PMS identification
unambiguously. Eventually, the first spectroscopically confirmed
discovery of a bona-fide T~Tauri star in the LMC, LTS J054427-692659,
a low-mass, late-type star located within the dark cloud Hodge II 139,
was reported by Wichmann, Schmitt \& Krautter~(\cite{wic01}).

Some of the evidence from ground and HST-based studies shows that
there may be significant differences between star formation processes
in the LMC and in the Galaxy. For example, Lamers et al~(\cite{lam99})
and de Wit et al~(\cite{wit02}), on the basis of ground-based data,
have suggested the presence of high-mass Pre-Main Sequence stars in
the LMC (Herbig AeBe stars, but see the caveats in de Wit et
al~\cite{wit05}) with luminosities systematically higher than observed
in our Galaxy, and located well above the ``birthline'' of Palla \&
Stahler~(\cite{ps91}). They attribute this finding either to a
shorter accretion timescale in the LMC or to its smaller dust-to-gas
ratio. Whether such differences in the physical conditions under which
stars form will generally lead to differences at the low mass end is
an open question, but Panagia et al~(\cite{pan00}, Paper I) and
Romaniello et al~(\cite{rom04}) offer tantalising evidence in this
direction.

HST-based studies of the IMF in the LMC (which have mostly
concentrated on young, compact clusters) have in fact produced, for
the lower masses, often discrepant results. On the one hand, there is
widespread agreement that the IMF for $M>3 M_{\sun}$ is similar to the
mean Galactic one (i.e. a ``Salpeter function'', a power-law with
slope $\alpha=-2.35$). On the other hand, though, the results for the
lower masses ($1 M_{\sun}\lesssim M\lesssim3 M_{\sun}$) are rather
difficult to reconcile with each other, with different authors finding
(in different regions, or even in the same one) very different slopes,
ranging from very steep IMFs ($\alpha\simeq-3$, Mateo~\cite{mat88}) to
quite shallow ones ($\alpha\simeq-1$, Sirianni et~al~\cite{sir00})
with many intermediate values represented.

As a part of an ongoing effort to characterise the young stellar
population in the LMC, we have applied the techniques developed by
Romaniello~(\cite{rom98}) to detect and characterise low mass Pre-Main
Sequence (PMS) stars in two regions of recent star formation observed
with the WFPC2 on board the HST. The fields were selected in the HST
archive so as to have deep narrow-band \ha\ observations, which, as we
shall see, are fundamental to detect the low mass Pre-Main Sequence
populations. Both fields are in the proximity of
\object{Supernova~1987A} (SN1987A), whose immediate vicinities we have
already studied with similar methods (Romaniello~\cite{rom98}, Panagia
et al \cite{pan00}, Romaniello et al~\cite{rom04}). In addition to
\ha, one of the two fields presented here was imaged in four broad
band filters, so that this wealth of data allowed us to recover the
intrinsic properties of the stars as well as their reddening following
the prescriptions developed by Romaniello et al~(\cite{rom02}). For
the other field the broad band data were limited to two filters only,
and, therefore, the analysis had to be necessarily cruder. We will,
however, show how interesting conclusions on the properties of the
stellar populations can be drawn also from such a limited
dataset. More fields are in the process of being analysed to enlarge
our study of young populations in the Magellanic Clouds across the
mass spectrum.

Section~\ref{sec:obs} is devoted to the description of the datasets
and the reduction techniques, whereas the results for the two fields
considered here are discussed in section~\ref{sec:south} and
\ref{sec:west}. Finally, summary and conclusions are presented in
section~\ref{sec:s&c}.

\section{Observations and data reduction\label{sec:obs}}
We have taken advantage of the vast HST archive to select two LMC
fields with recent star formation activity. They were chosen for
having both broad-band and \ha\ observations, which, as we shall see
in section~\ref{sec:south_tta}, is a fundamental requirement to
identify the low mass Pre-Main Sequence stars, and unambiguously
disentangle them from the much older field subgiants.

\begin{figure}[!ht]
\begin{center}
\vbox{\epsfig{figure=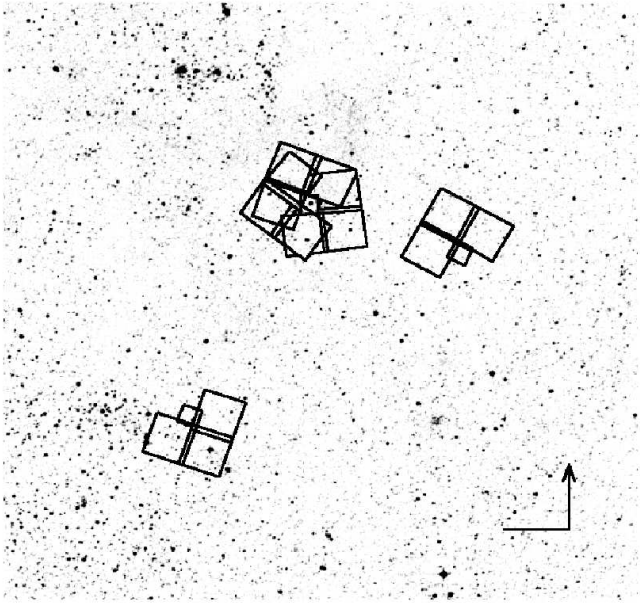,width=\linewidth}}
\end{center}
\caption[]{Footprints of the WFPC2 pointings superposed to a DSS image
$20\arcmin$ on the side.  The progenitor of SN1987A is visible at the
centre of the three WFPC2 fields of Panagia et
al~(\protect{\cite{pan00}}). The 30~Doradus nebula is roughly
$20\arcmin$ to the northeast.}
\label{fig:dss}
\end{figure}

The two fields we present in this paper are located some $20\arcmin$
southwest of the \object{30~Doradus} nebula. This area includes
regions of very active star formation, in which different groups of
early type stars are interspersed with HII regions and Supernova
remnant shells, as well as one of the best know objects in the entire
sky: Supernova~1987A. The footprints of the WFPC2 pointings
superimposed on a DSS image are shown in Figure~\ref{fig:dss}. Also
shown in the same Figure are the locations of the fields around
SN1987A itself in which Panagia et al~(\cite{pan00}) and and
Romaniello et al~(\cite{rom04}) have identified nearly 500 Pre-Main
Sequence candidates through their \ha\ and Balmer continuum emission.

The centres of the fields are located at $\alpha=05:36:11.57$,
$\delta=-69:23:00.74$ (J2000) and $\alpha=05:34:38.01$,
$\delta=-69:17:14.90$ (J2000). Given their position with respect to
the SN1987A, in the following we will refer to the pointings described
here as South field and West field, respectively. Their angular
separations from SN1987A are $8\arcmin$ and $4.5\arcmin$ (\ie 120 and
68~pc in projection), respectively, and the South field is just
$2.5\arcmin$ west of the young cluster \object{NGC~2050}. The logs of
the observations of the two fields are reported in
Table~\ref{tab:log_south} and~\ref{tab:log_west}. The detailed
description of the WFPC2 camera and its filters can be found in the
corresponding Instrument Handbook (Heyer et al~\cite{wfpc2}).

\begin{table}[!ht]
\begin{tabular}[b]{lcc}\hline
Filter Name & Exposure Time (s) &     Comments   \\ \hline
{\bf F300W} &    $600+1200$     &  Wide U \\
{\bf F450W} &     $50+200$      &  Wide B \\
{\bf F675W} &     $50+200$      &     R-like \\
{\bf F814W} &     $140+400$     &     I-like \\
{\bf F656N}  &    $1200+1200+1300$    &       \ha      \\ \hline
\end{tabular}
\caption[]{Log of the observations of the South field. They were taken
on January, $1^\mathrm{st}$ 1997, under proposal number 6437, PI Robert
P. Kirshner.}
\label{tab:log_south}
\end{table}

\begin{table}[!ht]
\begin{minipage}{\linewidth}
\begin{tabular}{lcc}\hline
Filter Name & Exposure Time (s) &     Comments   \\ \hline
{\bf F547M} & $2\times 120^{\mathrm{a}} $       & V-like\\
{\bf F675W} & $2\times 120^{\mathrm{a}}$       & R-like \\
{\bf F656N} & $1000+2\times2000^{\mathrm{a}}$  & \ha\\
            & $900+1000+2000^{\mathrm{b}}$     & \\\hline
\protect{\footnotetext[1]{June, $24^\mathrm{th} 1996$.}}
\protect{\footnotetext[2]{June, $25^\mathrm{th} 1996$.}}
\end{tabular}
\end{minipage}
\caption[]{Log of the observations of the West field. The proposal number
is 6033, PI Jeremy Walsh.}
\label{tab:log_west}
\end{table}

The data were processed through the standard Post Observation Data
Processing System pipeline for bias removal and flat fielding. 
In all cases the cosmic ray events were removed combining the
available images, duly registered.

The plate scale of the camera is 0.045 and 0.099 arcsec/pixel in the
PC and in the three WF chips, respectively. We performed aperture
photometry following the prescriptions by Gilmozzi~(\cite{gil90}) as
refined by Romaniello~(\cite{rom98}), \ie measuring the flux in a
circular aperture of 2~pixels radius and the sky background value in
an annulus of internal radius 3~pixels and width 2~pixels. Due to the
undersampling of the WFPC2 PSF, this prescription leads to a smaller
dispersion in the Colour-Magnitude Diagrams, \ie better photometry,
than PSF fitting for non-jittered observations of marginally crowded
fields (Cool \& King~\cite{coo95}, Romaniello~\cite{rom98}). The
photometric error is computed taking into account the poissonian
statistics of the flux from the source and the measured rms of the
flux from the sky background around it. This latter includes the
effects of poissonian fluctuations on the background emission,
flat-fielding errors, etc.

Photometry for the saturated stars was recovered by either fitting
the unsaturated wings of the PSF for stars with no saturation outside
the central 2 pixel radius, or by following the method developed by
Gilliland~(\cite{gil94}) for the heavily saturated ones.

The flux calibration was obtained using the internal calibration of the
WFPC2 (Whitmore~\cite{whit95}), which is typically accurate to within 5\% at
optical wavelengths. The spectrum of Vega is used to set the photometric
zeropoints (VEGAMAG system).

\section{The South field\label{sec:south}}
\begin{figure}[!ht]
\vbox{\psfig{figure=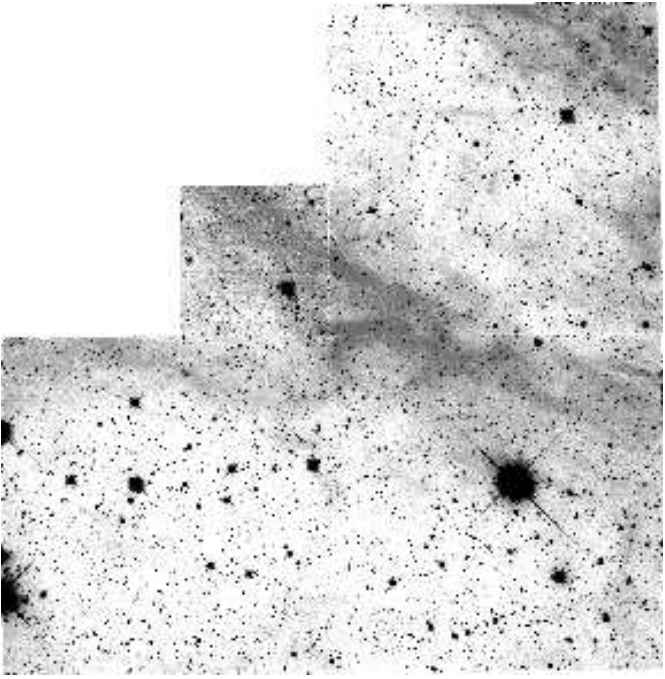,width=\linewidth}}
\caption[]{The South field as observed in the combination of F450W
(B), F675W (R) and F656N (\ha) filters.}
\label{fig:south_color}
\end{figure}

The F450W (B), F675W (R) and F656N (\ha) composite image of the South
field is shown in Figure~\ref{fig:south_color}.

We have detected 13,098 stars in the whole field, 4,108 of which with
a mean error in the 4 broad bands lower than 0.1~mag:


\begin{equation}
\bar{\delta}_4=\sqrt{\frac{\delta{_\mathrm{F300W}^2}+
\delta{_\mathrm{F450W}^2}+\delta{_\mathrm{F675W}^2}+
\delta{_\mathrm{F814W}^2}}{4}}<0.1
\label{eq:delta4}
\end{equation}

The brightest star in the field, located in the WF3 chip, is so badly
saturated that it was impossible to measure its magnitude from the
WFPC2 images. On the other hand, it is a well known star,
\object{Sk~-69~211} (Sanduleak~\cite{sand69}), also known as
\object{HDE~269832}, and its accurate photometry is available from
ground-based studies. Fitzpatrick~(\cite{fit88}) classifies it as B9Ia
supergiant with V$=10.36$, B$-$V$=0.09$ and $\ebv\simeq0.1$. With a
radial velocity of $+272~\mathrm{km~s}^{-1}$ it is a bona fide member
of the LMC (Ardeberg et al~\cite{ard72}). Using the temperature scale
and bolometric corrections of Schmidt-Kaler~(\cite{sk82}), one derives
for it an effective temperature of roughly 10,000~K and a luminosity
of roughly $3\cdot10^5$~L$_{\sun}$.

\begin{figure}[!ht]
\vbox{\psfig{figure=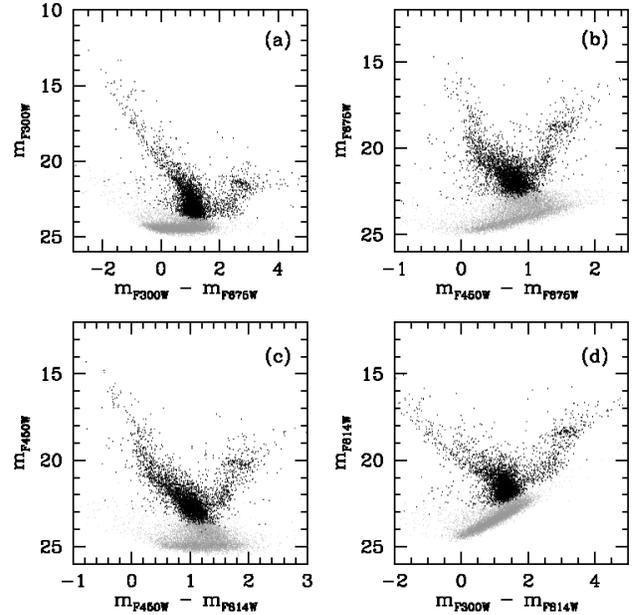,width=\linewidth}}
\caption[]{Colour-Magnitude diagrams of the 13,098 stars detected in
the South field for four combination of filters. The black dots are
the 4,108 stars with a mean photometric error in the four bands
($\bar{\delta}_4$, see equation~\ref{eq:delta4}) smaller than
0.1~mag. Sk~-69~211 is not shown.}
\label{fig:south_cmd}
\end{figure}

The Colour-Magnitude Diagrams for four combination of filters are shown in
Figure~\ref{fig:south_cmd}.

The presence of at least two distinct populations is readily detected
in Figure~\ref{fig:south_cmd}. The first one is associated with the
feature extending up to \mbox{$m_\mathrm{F450W}\simeq~15$} and with a
colour \mbox{$(m_\mathrm{F450W}-m_\mathrm{F814W})\simeq-0.6$} (see
panel~(c)). Of course, this is a young generation of stars still on or
just off the Zero Age Main-Sequence. The old population of the LMC
field, with ages in excess of 500~Myr or so, is identified by the
presence of the Red Clump at \mbox{$m_\mathrm{F450W}\simeq~20$},
\mbox{$(m_\mathrm{F450W}-m_\mathrm{F814W})\simeq~1.8$} and an extended
Red Giant Branch reaching magnitudes as bright as
\mbox{$m_\mathrm{F450W}\simeq~18$} at
\mbox{$(m_\mathrm{F450W}-m_\mathrm{F814W})\simeq~3$}.

\subsection{Reddening distribution\label{sec:south_ebv}}
In order to derive the detailed properties of the stars in the field
we first have to determine the interstellar reddening. As in any
region of recent star formation, one can expect the stars to be still
entwined with the material they were formed from and, hence, the
interstellar reddening to be spatially inhomogeneous. We have, then,
applied the procedure developed by Romaniello et al~(\cite{rom02}) to
measure \ebv, $\mathrm{T}_{eff}$ and luminosities for individual
stars. We have used a value for the distance to the LMC of 51.8~kpc
(Romaniello et al~\cite{rom00}). In brief, the procedure consists in a
fit of the observed fluxes to a grid of theoretical ones, reddened by
various amounts of \ebv. A minimum $\chi^2$ technique is, then, used
to identify the best combination of $\mathrm{T}_{eff}$, \ebv\ and
luminosities, with their associated errors. As inputs for the fit we
have used the atmosphere models by Bessell et al~(\cite{bes98}) and
the reddening law appropriate for this region of the LMC as derived by
Scuderi et al~(\cite{scud96}) from a study of \object{Star~2}, one of
the two stars projected within a few arcseconds from SN1987A.

We were able to estimate \ebv\ for 1,205 stars, \ie 9\% of the
total or one every 4 square arcseconds on average, while for the
others we have used the mean value of their neighbours with direct
measurements (cf. Romaniello et al~\cite{rom02}). An inspection
of Figure~\ref{fig:south_uvex}, which will be discussed in detail in
section~\ref{sec:south_tta_uv}, shows that the location of the
dereddened stars in the $(m_\mathrm{F300W,0}-m_\mathrm{F450W,0})$ vs
$(m_\mathrm{F450W,0}-m_\mathrm{F814W,0})$ colour-colour plane
generally agrees very well with the expectations from the Bessell et
al~(\cite{bes98}) model atmospheres, with a scatter consistent with
observational errors ($\bar{\delta}_4\lesssim0.1$, the discrepancy
observed for some stars at
$(m_\mathrm{F450W,0}-m_\mathrm{F814W,0})\simeq0.7$ is thoroughly
discussed in section~\ref{sec:south_tta_uv}). This agreement, then,
gives us confidence on the use of the Bessel et al~(\cite{bes98})
atmospheres to reproduce the observed colors and confirms that our
dereddening procedure does not introduce any significant systematic
effect (see also Romaniello et al~\cite{rom02}).

\begin{figure}[!ht]
\vbox{\psfig{figure=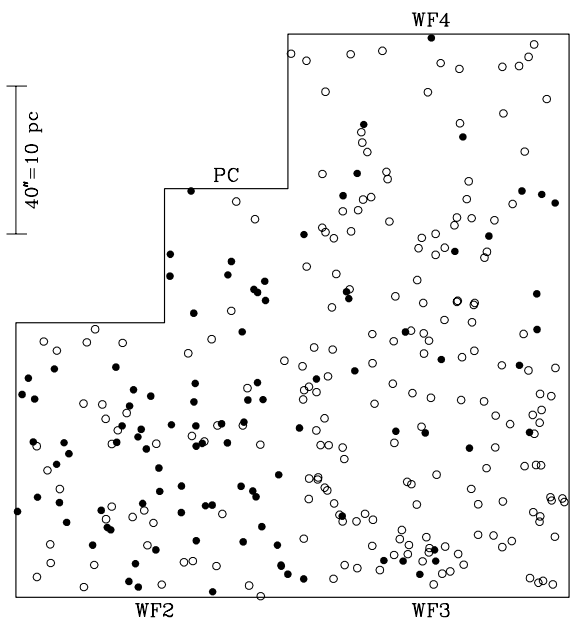,width=\linewidth}}
\caption[]{Spatial distribution of stars with high ($\ebv>0.3$,
filled circles) and low ($\ebv<0.1$, open circles) reddening. The
average reddening is significantly higher in the WF2 chip, \ie towards
NGC~2050 (see Figure~\ref{fig:dss}).}
\label{fig:south_ebv}
\end{figure}

The derived reddening distribution is indeed rather clumpy, as shown
in Figure~\ref{fig:south_ebv}.  The mean extinction over the field is
$\ebv=0.180$ (second column of Table~\ref{tab:south_ebv}), \ie
comparable to the mode of the distribution for the old population of
the LMC as a whole of $\ebv\simeq0.1$ (Zaritsky~\cite{zar99}). The
corresponding rms is 0.091 (third column). Thanks to the large number
of stars, then, the error on the mean is negligible, \ie
$0.091/\sqrt(1205)=0.003$. Again from Table~\ref{tab:south_ebv}, one
sees that the mean formal error on the individual reddening
determination due to the combined effects of photometric errors and
fitting uncertainties (fourth column) is 0.029~magnitudes. This,
together with the measured rms, implies that the intrinsic dispersion
in \ebv\ due to the patchy distribution of dust is
0.086~magnitudes. This value is both statistically significant
and non negligible, considering that it reflects itself in a spread of
about $0.086\times5.5\simeq0.5$~mag in F300W,
$0.086\times3.9\simeq0.3$~mag in F450W, $0.086\times2.4\simeq0.2$~mag
F675W and $0.086\times1.9\simeq0.15$~mag in F814W (see Table~8 of
Romaniello et al \cite{rom02} for the extinction coefficients in the
WFPC2 bands).

It is also clear from Figure~\ref{fig:south_ebv} that the interstellar
extinction is inhomogeneous on larger scales as well, with the
reddening being significantly higher in the PC and WF2 chips, \ie in
the direction of NGC~2050, than elsewhere. The actual numbers are
reported in Table~\ref{tab:south_ebv}, where one can see that the
difference in the mean reddening between the PC and WF2 chips on the
one side and the WF3 and WF4 chips on the other is statistically
highly significant.  The dispersion, on the other hand, is almost
constant across the chips (see also section~\ref{sec:south_space}).
All of these facts highlight once more how crucial it is to accurately
deredden the stars in order to recover and interpret their properties.

\begin{table}[!ht]
\begin{tabular}[b]{*{6}{c}}\hline
Chip & Mean & rms & Error & Dispersion & No. of stars\\\hline
{\bf PC}  & 0.233 & 0.117 & 0.037 & 0.112 & 33\\
{\bf WF2} & 0.210 & 0.093 & 0.027 & 0.089 & 401\\
{\bf WF3} & 0.157 & 0.089 & 0.031 & 0.083 & 427\\
{\bf WF4} & 0.168 & 0.077 & 0.028 & 0.072 & 344\\\hline
     All  & 0.180 & 0.091 & 0.029 & 0.086 & 1,205
\end{tabular}
\caption[]{Spatial distribution of \ebv\ colour excess as measured
in the South field. In column~3 we report the measured rms of the \ebv\
distribution, in column~4 the mean error on \ebv\ resulting from the
fitting procedure (see text), whereas column~5 lists the intrinsic reddening
dispersion, computed as the quadratic difference of the first two quantities.
The number of stars with direct extinction determination is shown in the
last column.}
\label{tab:south_ebv}
\end{table}

Finally, it is interesting to notice that the spatial distribution of
\ebv\ does not correlate with the diffuse \ha\ emission, which is
highest across the PC and the upper part of the WF3 chips, whereas it
is quite faint in the WF2 chip (see Figure~\ref{fig:south_color}).

\subsection{HR diagram\label{sec:south_hr}}
\begin{figure}[!ht]
\vbox{\psfig{figure=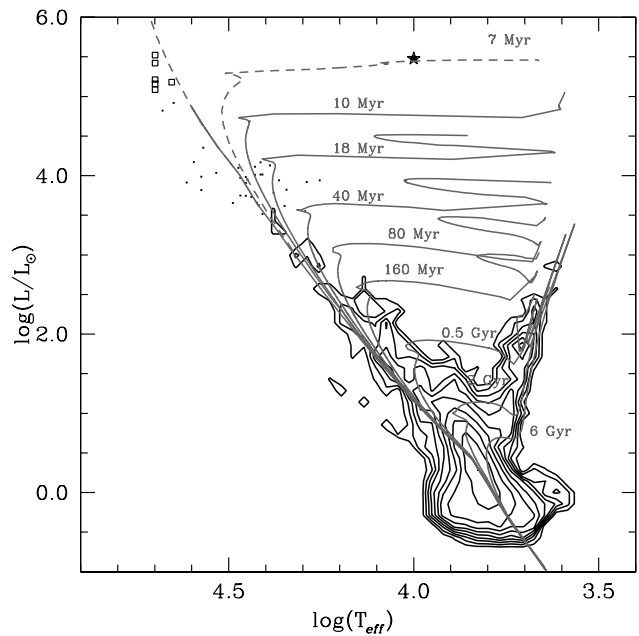,width=\linewidth}}
\caption[]{Hess diagram for the 13,098 stars in the South field.
The open squares indicate 6 stars for which the F300W magnitude is ill
determined because of saturation and, hence, the fit was performed
excluding this filter. All of them, but one, require a temperature
equal to or larger than 50,000~K, the highest available for the model
atmospheres by Bessell et al~(\protect{\cite{bes98}}). The location of
\mbox{Sk~-69~211} according to the photometry by
Fitzpatrick~(\protect{\cite{fit88}}) is shown with a star symbol.  The
ZAMS and selected isochrones for Z$=0.3\cdot\mathrm{Z}_{\sun}$ by
Brocato \& Castellani~(\protect{\cite{bc93}}) and Cassisi et
al~(\protect{\cite{ccs94}}) are overlaid to the data (grey solid
lines). These models only extend to $25~M_{\sun}$, so we have
complemented them with those of Schaerer et al~(\cite{sch93}) for
higher masses (grey dashed lines).}
\label{fig:south_hr}
\end{figure}

Once dereddened, the stars can accurately be placed in the HR diagram
and their properties derived by comparing their positions with theoretical
models (Figure~\ref{fig:south_hr}).

As we have already noticed from an inspection of the CMDs shown in
Figure~\ref{fig:south_cmd}, no single age can explain the distribution
of stars as observed in the HR diagram of the South field. Starting
from the top-left corner of the HR diagram, the most massive stars,
plotted as open squares in Figure~\ref{fig:south_hr}, have masses of
the order of $40~M_{\sun}$ and, thus, are definitely younger than
5~Myr. Since they are severely affected by saturation in both F300W
exposures, their intrinsic parameters were derived excluding the flux
in this filter from the fit. As discussed by Romaniello et
al~(\cite{rom02}), lacking the UV information, the fit is rather
uncertain. In any case, all of them, but one, seem to require
temperatures higher than 50,000~K, the highest available for the
Bessell et al~(\cite{bes98}) models. In conclusion, although their
stellar parameters cannot be accurately measured, these stars are
certainly massive and, hence, young, even more so than their position
in the HR diagram would indicate.

With a comparable bolometric luminosity, but a much lower temperature,
there is Sk~-69~211. It falls exactly on the 7~Myr isochrone by
Schaerer et al~(\cite{sch93}), \ie it is without a doubt older than
the bluest stars in the field. A comparison with stellar evolutionary
models assigns a mass of roughly $27~\mathrm{M}_{\sun}$ to
it. Interestingly, a similar phenomenon was reported in Galactic OB
associations by Massey et al~(\cite{mas95}), who noted the
``occasional presence of an evolved star'' among a younger population.

Finally, stars with ages between a few tens of million and several
billion years are required to account for the stars at lower
luminosities.

\subsection{A population of peculiar objects\label{sec:south_pec}}
A population of peculiar objects was detected in the South field by
means of their \ha\ and Balmer continuum excesses (the two emissions
positively correlate, as we shall see in
section~\ref{sec:south_tta_uv}). In this section we describe the
selection criteria for these stars and offer a possible explanation of
their nature as Pre-Main Sequence stars.

\subsubsection{\ha\ emission\label{sec:south_tta_ha}}
Following Romaniello~(\cite{rom98}; see also Panagia et
al~\cite{pan00} and Romaniello et al~\cite{rom04}), we compare the
magnitudes in the R band (F675W) with the ones measured with the
narrow band \ha\ filter (F656N) to identify stars with a sufficiently
strong \ha\ emission. As it can be seen in
Figure~\ref{fig:south_color} there is significant nebular \ha\
emission in the South field. It is distributed in filaments with a
typical length of several tens of arcseconds up to several
arcminutes. All of the stars that are found to have \ha\ emission (see
below) were inspected by eye to ensure that the nebular contribution
was properly subtracted from the narrow band photometry.

In order to select stars with a statistically significant \ha\
emission we have compared the observed histogram of
$(m_\mathrm{F675W}-m_\mathrm{F656N})$ for the stars in the field with
the distribution expected if no emission was present. This latter was
derived with the following four steps:

\begin{figure}[!ht]
\vbox{\psfig{figure=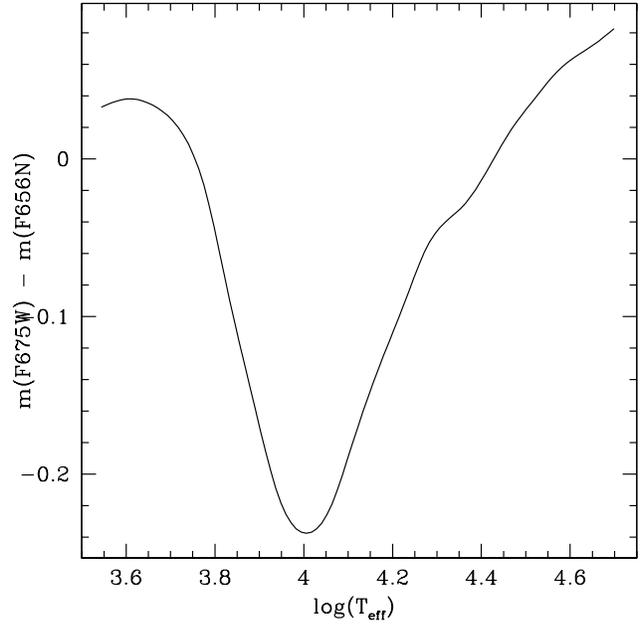,width=\linewidth}}
\caption[]{Relation between the WFPC2
$(m_\mathrm{F675W}-m_\mathrm{F656N})$ colour (\ie $\mathrm{R}-\ha$)
and effective temperature as derived from Kurucz~(\cite{kur93}) model
atmospheres (see text).  }
\label{fig:tedm}
\end{figure}

\begin{figure}[!ht]
\vbox{\psfig{figure=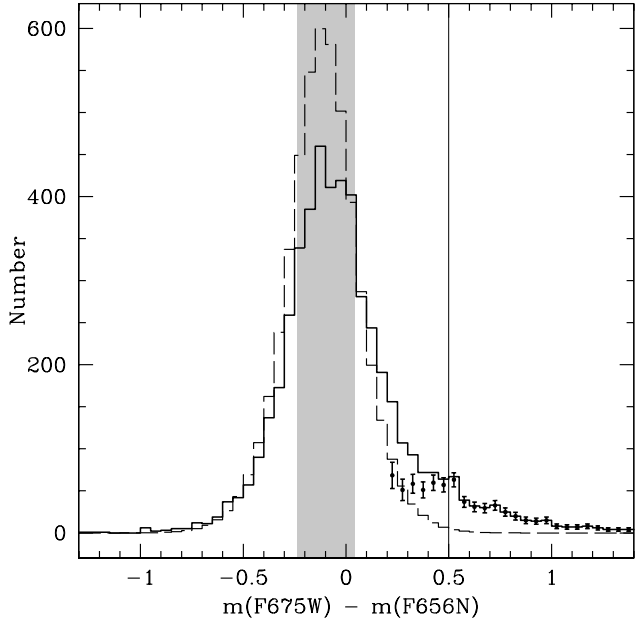,width=\linewidth}}
\caption[]{Observed (thick solid line) and expected (thin dashed line,
see text) distribution of $(m_\mathrm{F675W}-m_\mathrm{F656N})$ colour
in the South field. The histograms contain 4,938 stars, those with a
positive detection in the F656N (\ha). The shaded areas marks the
intrinsic colour interval predicted by the atmosphere models, \ie
without the effect of observational errors (see also
Figure~\ref{fig:tedm}).  As it can be seen, the left-hand tails of the
two histograms are in agreement, and the excess of observed stars with
positive $(m_\mathrm{F675W}-m_\mathrm{F656N})$ colours, \ie \ha\
emission, compared to the expectations is evident for
$(m_\mathrm{F675W}-m_\mathrm{F656N})\gtrsim0.2$. The distribution of the
stars with significant excess (see text) is shown as dots, together with
the corresponding errorbars. The value of the excess above which stars
can be identified individually is marked with a vertical line.}
\label{fig:south_ha-ex_all}
\end{figure}

\begin{enumerate}
  \item Kurucz~(\cite{kur93}) model atmospheres for a metallicity
$[\mathrm{M}/\mathrm{H}]=-0.5$ at different effective temperatures
were used to compute synthetic spectra between 5,700 and 7,800~\AA,
the region covered by the F675W filter. A spectral resolution of
1~\AA\ was chosen, so as to sample the \ha\ line to a much better
accuracy than the FWHM of the WFPC2 F656N filter ($\sim20$~\AA, Heyer
et al~\cite{wfpc2}).
  \item These spectra were fed to the \texttt{synphot} task in IRAF to
yield the expected $(m_\mathrm{F675W}-m_\mathrm{F656N})$ colour of
``normal'' stars, \ie with \ha\ in absorption, as a function of their
effective temperature. The resulting relation is shown in
Figure~\ref{fig:tedm}.
  \item We have used this relation together with the measure effective
temperatures of the stars (cfr section~\ref{sec:south_hr}) to compute
the expected $(m_\mathrm{F675W}-m_\mathrm{F656N})$ colour for our
stars in the absence of any \ha\ emission (shaded area in
Figure~\ref{fig:south_ha-ex_all}).
  \item Finally, this intrinsic distribution was broadened by the
measured photometric error on the
$(m_\mathrm{F675W}-m_\mathrm{F656N})$ colour.  To this end, the
intrinsic distribution was convolved with the appropriate error
function, \emph{i.e.} the normalised sum of the error Gaussians of the
individual stars. This is, then, the colour distribution that the
stars in the field are expected to have according to their measured
temperature.
\end{enumerate}

The comparison of the observed and expected distributions derived as
described above is presented in Figure~\ref{fig:south_ha-ex_all}
(solid and dashed lines, respectively). The left-hand tails of the two
distributions are in good agreement with each other, implying that the
broadening towards negative values of
$(m_\mathrm{F675W}-m_\mathrm{F656N})$ compared to the intrinsic,
error-free distribution (shaded area in
Figure~\ref{fig:south_ha-ex_all}) is caused by photometric
errors. Moving towards more positive
$(m_\mathrm{F675W}-m_\mathrm{F656N})$ values, however, the two
histograms become increasingly discrepant, until an excess of observed
stars with positive $(m_\mathrm{F675W}-m_\mathrm{F656N})$ colours, \ie
\ha\ emission, compared to the expectations is evident for
$(m_\mathrm{F675W}-m_\mathrm{F656N})\gtrsim0.2$ (dots in
Figure~\ref{fig:south_ha-ex_all}). Applying a $\chi^2$ test confirms
that the distribution are conclusively different
($\tilde{\chi}^2=11$).

\begin{figure}[!ht]
\vbox{\psfig{figure=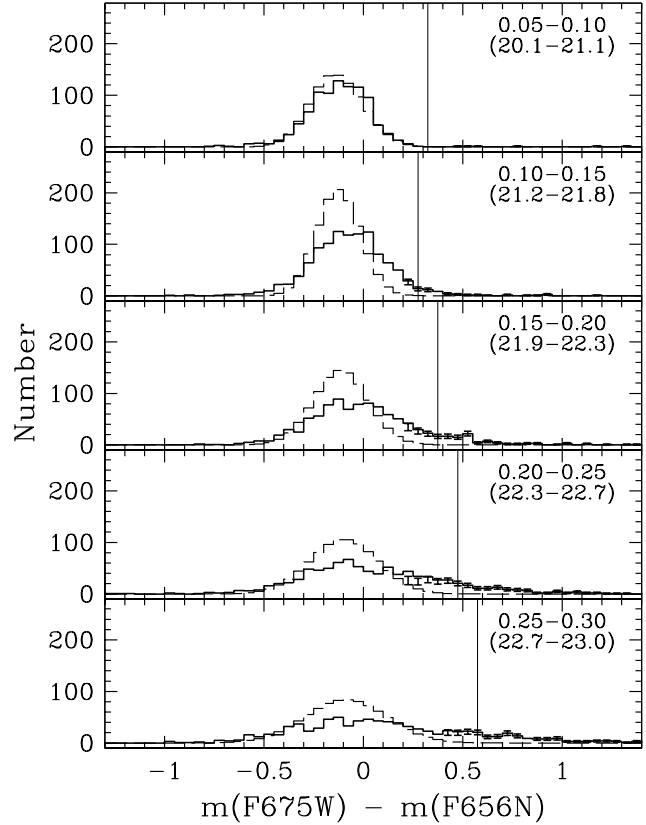,width=\linewidth}}
\caption[]{Same as Figure~\ref{fig:south_ha-ex_all}, but each panel
only contains the stars in the quoted range of error on the
$m_\mathrm{F675W}-m_\mathrm{F656N}$ colour. The corresponding
approximate range in $m_\mathrm{F675W}$ magnitude is also reported in
each panel. The value of the excess above which stars can be
identified individually is marked with a vertical line in each
panel.}
\label{fig:south_ha-ex_five}
\end{figure}

In Figure~\ref{fig:south_ha-ex_five} we report the observed and
expected $(m_\mathrm{F675W}-m_\mathrm{F656N})$ distributions, derived
as described above, broken down in different intervals of
observational uncertainty
$\delta(m_\mathrm{F675W}-m_\mathrm{F656N})$. The error range used is
quoted in each panel. Since the photometric error is dominated by
Poisson noise on photon statistics, the intervals in error are
essentially intervals in $m_\mathrm{F675W}$ magnitude (reported within
brackets in each panel). In all cases the widths of the left-hand side
of the observed and expected distributions agree, confirming that the
broadening is, indeed, due to photometric errors, whereas an excess of
\ha-emitting stars is detected at
$(m_\mathrm{F675W}-m_\mathrm{F656N})\gtrsim0.2$.

Based on the distributions of Figure~\ref{fig:south_ha-ex_all}
and~\ref{fig:south_ha-ex_five}, we use the following criteria to
identify the stars with a statistically significant excess:

\begin{itemize}
  \item the stars that populate $(m_\mathrm{F675W}-m_\mathrm{F656N})$ bins 
in which the observed distribution exceeds the expected one by 3 Poissonian
standard deviations or more;
  \item while the criterion above ensures a high statistical
significance of the selected objects, by definition it excludes
stars in scarcely populated bins, \ie with less than 9 stars. To
compensate for this, we also include in the sample the stars for which
the $(m_\mathrm{F675W}-m_\mathrm{F656N})$ excess itself is highly
significant, \ie $(m_\mathrm{F675W}-m_\mathrm{F656N})>
3\cdot\delta(m_\mathrm{F675W}-m_\mathrm{F656N})$, even if they belong
to sparse bins. This only adds roughly 1\% of the stars to the sample,
but they are important ones because they have the highest \ha\
emission and, hence, are the most active ones.
\end{itemize}

In total, there are $781\pm33$ stars that satisfy the criteria
above. Their distribution, with the associated error bars, is shown in
Figure~\ref{fig:south_ha-ex_all} as dots. These errors are the
statistical ones that pertain to this particular dataset. Deeper
observations, most notably in the \ha\ F656N filter, would lead to a
higher number of detections (see the discussion in
section~\ref{sec:west_tta}).

The number of excess stars in each panel of
Figure~\ref{fig:south_ha-ex_five} is reported in
Table~\ref{tab:south_ha-ex_five}, together with the ratio of the total
number of stars to that of the excess stars, from where one can see
that roughly 40\% of the 781 \ha\ emitters have quite good photometry,
\ie an error smaller than 0.2~mag. This generally implies errors in
both bands less than 0.14, \ie a 7 sigma detection.

\begin{table}[!ht]
\begin{tabular}[b]{lcc}\hline
Error range &  Number of   &      Ratio     \\
            & excess stars & (total/excess)\\\hline
$0.05-0.10$ &   $7\pm 3$ & 135 \\
$0.10-0.15$ &  $74\pm9$  &  16 \\
$0.15-0.20$ & $205\pm15$ &   5 \\  
$0.20-0.25$ & $264\pm18$ &   4 \\
$0.25-0.30$ & $231\pm15$ &   4 \\\hline
\end{tabular}
\caption[]{Distribution of \ha\ excess stars in different bins of
uncertainty on the \ha\ color index
$m_\mathrm{F675W}-m_\mathrm{F656N}$ (see text and
Figure~\ref{fig:south_ha-ex_five}).}
\label{tab:south_ha-ex_five}
\end{table}

It is clear from the discussion above that the objects with \ha\
excess can be divided in two groups, according the the strength of the
$(m_\mathrm{F675W}-m_\mathrm{F656N})$ excess. If it is so strong that
no ``normal'' star is expected to match it, \ie if the dashed
histogram in Figure~\ref{fig:south_ha-ex_all} is 0, then the stars can
be pinpointed one by one. As it can be seen in
Figure~\ref{fig:south_ha-ex_all} the condition that the expected
histogram be 0 corresponds to
$(m_\mathrm{F675W}-m_\mathrm{F656N})\gtrsim0.5$ (or
EW$(\ha)\gtrsim17$~\AA, see appendix~\ref{sec:app_ew-dm}). There are
366 such stars out of the total 781 with \ha\ excess (47\%). If, on
the other hand, the expected number of stars (dashed histogram in
Figure~\ref{fig:south_ha-ex_all}) is not zero (\ie
$m_\mathrm{F675W}-m_\mathrm{F656N}\lesssim0.5$), excess stars can only
be identified statistically, but not on a one-to-one basis.

\begin{figure}[!ht]
\vbox{\psfig{figure=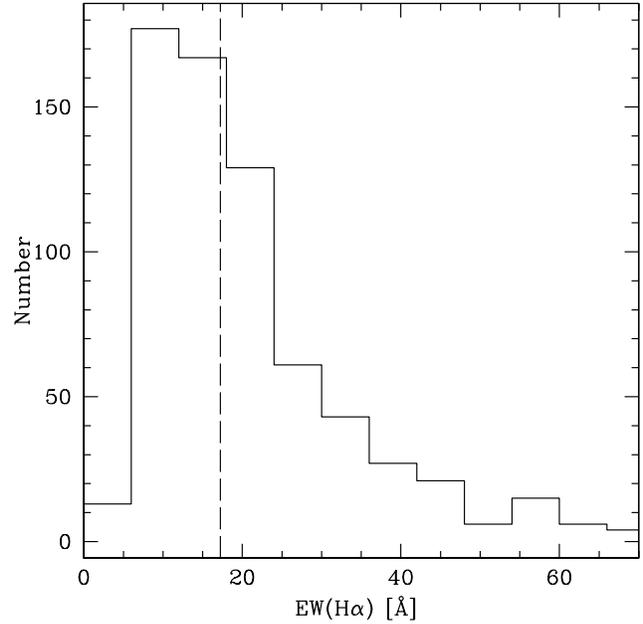,width=\linewidth}}
\caption[]{Distribution of \ha\ equivalent widths for the 781 stars
with statistically significant excess.  The observed colour excesses
$(m_\mathrm{F675W}-m_\mathrm{F656N})$ were converted to equivalent
widths using the relation derived in appendix~\ref{sec:app_ew-dm} and
displayed in Figure~\ref{fig:dmag_ew}. The dashed vertical line marks
the excess above which we can identify the emission-line stars
unambiguously ($m_\mathrm{F675W}-m_\mathrm{F656N}>0.5$, corresponding
to EW$(\ha)>17$~\AA).}
\label{fig:south_ew}
\end{figure}

The distribution of the \ha\ equivalent widths (EW) for the 781 stars
with statistically significant excess is displayed in
Figure~\ref{fig:south_ew}. The observed colour excesses
$(m_\mathrm{F675W}-m_\mathrm{F656N})$ were converted into to
equivalent widths using the relation derived in
appendix~\ref{sec:app_ew-dm} and plotted in
Figure~\ref{fig:dmag_ew}. The sharp drop at low values of the
equivalent width is not real, but, rather, an artifact of our
selection criteria that can only identify emission line objects above
$(m_\mathrm{F675W}-m_\mathrm{F656N})\gtrsim0.2$, \ie
$\mathrm{EW}(\ha)\gtrsim6$~\AA, the weaker ones being lost in the
noise.

\subsubsection{Balmer continuum excess\label{sec:south_tta_uv}}
Objects in the South field exhibit yet another spectral peculiarity in
addition to \ha\ emission: excess continuum emission blueward of the
Balmer jump at 3646~\AA, \ie between the F300W and F450W WFPC2
filters.

\begin{figure}[!ht]
\vbox{\psfig{figure=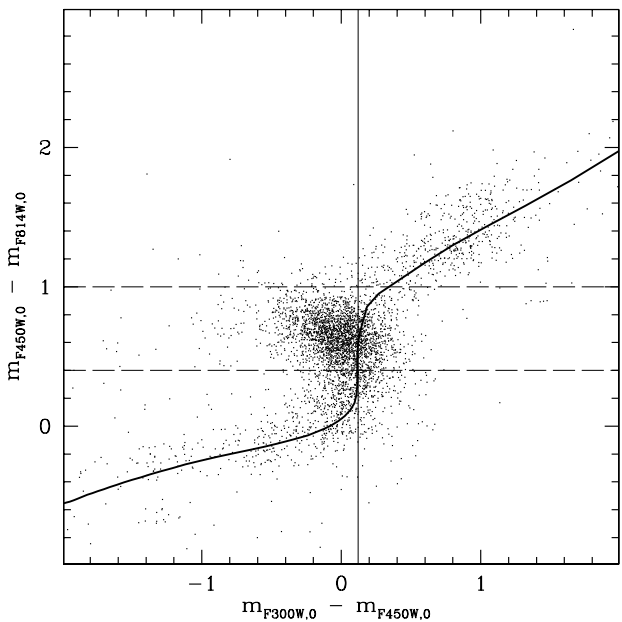,width=\linewidth}}
\caption[]{Dereddened $(m_\mathrm{F300W,0}-m_\mathrm{F450W,0})$ vs
$(m_\mathrm{F450W,0}-m_\mathrm{F814W,0})$ colour-colour diagram for the
stars in the South field with a mean error in the 4 wide bands
($\bar{\delta}_4$, see equation~\ref{eq:delta4}) smaller than 0.1. The
theoretical locus from the stellar atmosphere models of Bessell et
al~(\protect{\cite{bes98}}) is shown as a solid line and the expected
$(m_\mathrm{F300W,0}-m_\mathrm{F450W,0})$ colour at
$(m_\mathrm{F450W,0}-m_\mathrm{F814W,0})=0.7$ is marked by the thin
vertical line. The horizontal dashed lines highlight the region used
to select stars in Figure~\protect{\ref{fig:south_uv_ha}}.}
\label{fig:south_uvex}
\end{figure}

The occurrence of this excess emission can be readily seen in
Figure~\ref{fig:south_uvex} where we plot the dereddened
$(m_\mathrm{F300W,0}-m_\mathrm{F450W,0})$ vs
$(m_\mathrm{F450W,0}-m_\mathrm{F814W,0})$ colours for stars with
overall good photometry ($\bar{\delta}_4<0.1$, see
equation~\ref{eq:delta4}). For reference, the expected locus from the
stellar atmosphere models of Bessell et al~(\cite{bes98}) for
$Z=0.3\cdot Z_{\sun}$ and $\log(g)=4.5$ is shown as a solid line.
As we have already noticed in section~\ref{sec:south_ebv}, there
is a general very good agreement between the locus of the dereddened
stars and the expectations from the models atmospheres, with a scatter
consistent with observational errors.  Thus, the \emph{asymmetric}
broadening compared to the models at
$(m_\mathrm{F450W,0}-m_\mathrm{F814W,0})\simeq0.7$ cannot be due to
photometric errors (in fact, the broadening towards redder
$(m_\mathrm{F300W,0}-m_\mathrm{F450W,0})$ colours is consistent with
the errors, the one to the blue is not). Rather, it has to be
ascribed to genuine deviations in the objects' spectra relative to a
bare photosphere, as represented by the solid line.

\begin{figure}[!ht]
\vbox{\psfig{figure=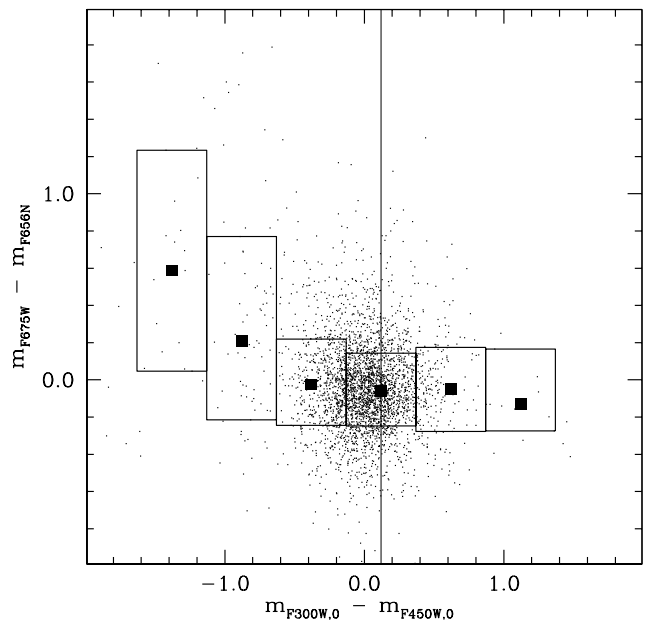,width=\linewidth}}
\caption[]{\ha\ excess vs Balmer continuum for stars in the South
field with $0.4<(m_\mathrm{F450W,0}-m_\mathrm{F814W,0})<1$ (black
dots, see Figure~\ref{fig:south_uvex}). The thin vertical line marks
the expected $(m_\mathrm{F300W,0}-m_\mathrm{F450W,0})$ colour in this
$(m_\mathrm{F450W,0}-m_\mathrm{F814W,0})$ interval from the models of
Bessell et al~\cite{bes98}. The filled squares represent the median
$(m_\mathrm{F675W}-m_\mathrm{F656N})$ value in the
$(m_\mathrm{F300W,0}-m_\mathrm{F450W,0})$ bin marked by the
rectangles, whose vertical extent includes 66\% of the stars in each
bin. The correlation between these two quantities is apparent and is
confirmed by the high value of Spearman's coefficient $\rho=-4.5$ (see
text).}
\label{fig:south_uv_ha}
\end{figure}

The excess Balmer continuum emission measured by the
$(m_\mathrm{F300W,0}-m_\mathrm{F450W,0})$ colour positively correlates
with the \ha\ excess, as shown in Figure~\ref{fig:south_uv_ha} where
we plot these two quantities against each another. The high
statistical significance of this correlation is confirmed by
performing Spearman's test on the data (e.g.,
Conover~\cite{con80}). The value of the correlation coefficient
$\rho=-4.5$ implies a probability of less than $7\cdot10^{-4}$ that
the two variables are uncorrelated. The fact that $\rho$ is negative
means that $(m_\mathrm{F675W}-m_\mathrm{F656N})$ increases as
$(m_\mathrm{F300W,0}-m_\mathrm{F450W,0})$ decreases towards more
negative values, \ie the equivalent width of the \ha\ emission
increases together with the Balmer continuum excess. Whereas the
quality of the data does not allow to derive the shape of the
correlation, its existence is proven with a high statistical
significance.

The fact that the \ha\ and Balmer continuum excesses correlate hints
at a common origin of the two phenomena.

\subsection{Are the peculiar objects Pre-Main Sequence stars?
\label{sec:south_tta}}
To recapitulate, in sections~\ref{sec:south_tta_ha} and
\ref{sec:south_tta_uv} we have described the detection in the South
field of a population of stars that exhibit \ha\ and Balmer continuum
excesses. Also, the fact that they correlate with each other points
towards a common origin for them. The question, then, arises as to
what they are and what causes their observed spectral peculiarities.

According to the histogram in Figure~\ref{fig:south_ew}, these stars
have \ha\ in emission with equivalent widths in excess of 6~\AA.
This, then, excludes any significant contamination of the sample by
stars with chromospheric activity, whose equivalent widths are smaller
than 3~\AA\ (e.g. Frasca \& Catalano~\cite{fra94}, White \&
Basri~\cite{whi03}).

Instead, we propose that the peculiar objects described in
section~\ref{sec:south_pec} are Pre-Main Sequence stars. This is based
on the following considerations:

\begin{enumerate}
\item As we have seen in section~\ref{sec:south_hr}, there are very
  young, very massive stars in the South field and it is only natural to
  expect the presence of equally young, but lower mass stars. In fact,
  Pre-Main Sequence evolutionary models indicate that stars less massive
  than about $2.5~\mathrm{M}_{\sun}$ are still approaching the Main
  Sequence after 5~Myr of evolution and occupy the same region of the
  HR diagram as the field population subgiants, which are evolving off
  the Main Sequence after several hundred millions or billions of years.

\item A class of Galactic Pre-Main Sequence stars, the Classical TTauri
  stars, are observed to have \ha\ emission with equivalent widths up to
  several hundred {\AA}ngstroms (e.g. Fern\'andez et
  al~\cite{fer95}). Moreover, both the \ha\ and Balmer continuum
  excesses in Pre-Main Sequence stars are thought to be linked to the
  mass accretion process from the circumstellar disk (e.g. Calvet et
  al~\cite{cal02}). As such, a correlation like the one we find between
  the two quantities, and which is illustrated in
  Figure~\ref{fig:south_uv_ha}, is to be expected.
\end{enumerate}

We refer the reader to Panagia et al~(\cite{pan00}) and Romaniello et
al~(\cite{rom04}) for an analogous population in the field of
SN~1987A, which is less than $10\arcmin$ away from the South field we
consider here (see Figure~\ref{fig:dss}).

\begin{figure}[!ht]
\vbox{\psfig{figure=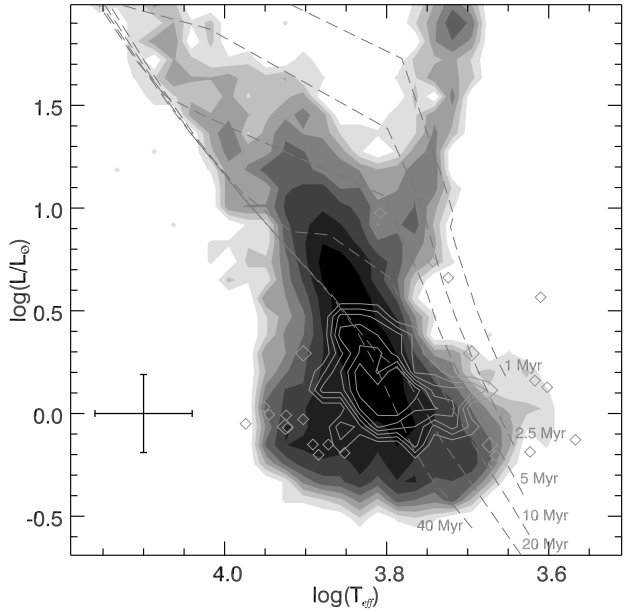,width=1.04\linewidth}}
\caption[]{Hess diagram for the general population in the South field
(greyscale) and the the 366 stars with unambiguous \ha\ emission in
the South field ($(m_\mathrm{F675W}-m_\mathrm{F656N})>0.5$, contours
and diamond symbols to indicate the position of individual
emitters).  Luminosity and temperature for the emission-line objects
are computed excluding the magnitude in the F300W filter, as it can be
contaminated by non-photospheric emission. Pre-Main Sequence
isochrones for various ages as computed by Siess et
al~(\protect{\cite{siess97}}) and the typical errorbar at the
luminosity of the emission-line stars are also shown.}
\label{fig:south_pms-hr}
\end{figure}

The positions in the HR diagram of the 366 \ha\ excess stars that can
be identified unambiguously
($(m_\mathrm{F675W}-m_\mathrm{F656N})>0.5$) are shown in
Figure~\ref{fig:south_pms-hr} together with the evolutionary tracks of
Siess et al~(\cite{siess97}). Luminosity and temperature for these
stars are computed excluding the magnitude in the F300W filter, so as
to avoid possible contamination from non-photospheric
emission\footnote{For about 75\% of the \ha\ emission stars the
effective temperatures determined with or without the F300W band
differ by less than 5\%. For the remaining 25\% of the objects, though,
the mean difference in $\mathrm{T}_{eff}$ is 20\%, in the sense that the
fit including the F300W band returns higher values, with a maximum
discrepancy of about 40\% for the stars with the strongest Balmer continuum
excesses.}. For comparison, there are roughly 7,500 stars in the same
area of the HR diagram. As we will discuss in
section~\ref{sec:south_mf} some of them might be Pre-Main Sequence
stars with an \ha\ emission that is too weak to be detected in the
available images. Most of them, however, belong to the much older LMC
population, and the 781 Pre-Main Sequence stars we can identify in a
statistical sense are outnumbered by a factor of about 10.

The comparison with Pre-Main Sequence isochrones by Siess et
al~(\cite{siess97}) shown in Figure~\ref{fig:south_pms-hr} confirms
the presence of a generation of young stars also among the low mass
stars (the evolutionary tracks indicate masses between 0.8 and 2
$M_{\sun}$ for them). The brightest and coolest among them could be as
young as 1-2~Myr, in agreement with what the upper Main Sequence
indicates.

\begin{figure}[!ht]
\vbox{\psfig{figure=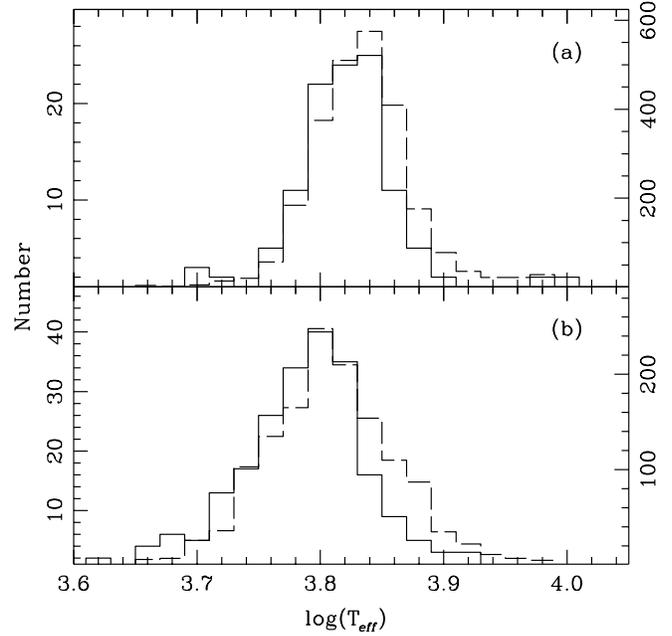,width=\linewidth}}
\caption[]{Temperature distribution of the stars for which the
\ha\ emission can be identified without ambiguity (solid line)
compared to the one of all of the stars (dashed line) in two intervals
of luminosity (see Figure~\ref{fig:south_pms-hr}):
$-0.2\lesssim\log(L/L_{\sun})\lesssim0.15$ (panel (a)) and
$0.15\lesssim\log(L/L_{\sun})\lesssim0.5$ (panel (b)). The
Kolmogorov-Smirnov test on the unbinned data confirms that the
distributions in each panel are different, with the one for the \ha\
emitters skewed towards lower temperature, as expected for Pre-Main
Sequence stars.}
\label{fig:south_te_his}
\end{figure}

However, the majority of the emission objects seem to have ages in
excess of 10~Myr.  A fraction of the emission objects even falls below
the locus of the Zero Age Main Sequence, a fact that, if taken at face
value, would be incompatible with their suggested Pre-Main Sequence
nature. Nonetheless, let us note that: \emph{(a)} the observed spread
in the HR diagram is consistent with being due to photometric and
dereddening errors (cfr the errorbar in
Figure~\ref{fig:south_pms-hr}), and \emph{(b)} the stars with \ha\
excess are, on average, colder than the stars at comparable
luminosity, as expected for Pre-Main Sequence objects. This is shown
in Figure~\ref{fig:south_te_his}, where the temperature distribution
for the candidate Pre-Main Sequence stars (solid line) is compared to
the one of all of the stars (dashed line) in two intervals of
luminosity (see Figure~\ref{fig:south_pms-hr}):
$-0.2\lesssim\log(L/L_{\sun})\lesssim0.15$ (panel (a)) and
$0.15\lesssim\log(L/L_{\sun})\lesssim0.5$ (panel (b)). Applying the
Kolmogorov-Smirnov (KS) test on the unbinned data and the $\chi^2$
statistics to the binned data confirm that the distribution are
conclusively different. In fact, the KS probability that the \ha\ and
non-\ha\ emitters are drawn from the same parent distribution is less
than 1.8\% for the brighter sample (panel (a)) and less than 0.04\%
for the fainter one (panel (b)). Similarly, the deviations of the
binned data correspond to $\chi^2$ values of 16.7 and 36.5,
respectively.

Taken at face value, the old ages inferred for the candidate Pre-Main
Sequence stars are at odds with the age of the brightest stars on the
upper Main Sequence stars ($\sim5$~Myr, see
section~\ref{sec:south_hr}). However, at this stage, we cannot claim
that this age difference is real and that star formation in this field
has proceeded on different timescale for different mass ranges. In
fact, a shift of about 10\% in temperature, in the sense of making the
candidate Pre-Main Sequence stars colder, is all that it is needed to
reconcile the two age determinations. Such a shift cannot be ruled out
on accounts of the uncertainties in determining the stellar parameters
from optical bands alone (see the discussion in Romaniello et
al~\cite{rom02}). Let us also note that a younger age for the
candidate Pre-Main Sequence stars would also help to reconcile them
with the current understanding of Galactic star-forming regions, where
evidences of accretion tend to disappear in most Pre-Main Sequence
stars after about 10~Myr (e.g. Calvet et al~\cite{cal05}, but see
Romaniello et al~\cite{rom04} for evidence in the LMC of vigorous
accretion at an age of about 14~Myr).

Converting the measured Balmer continuum excesses described in
section~\ref{sec:south_tta_uv} to a mass accretion rate following the
prescription of Gullbring et al~(\cite{gul98}, see also Robberto et
al~\cite{rob04}), one gets values from
several times $10^{-8}$ up to $10^{-7} M_{\sun}\ \mathrm{yr}^{-1}$. If
sustained for significantly longer than 10~Myr, accretion rates of
this magnitude imply an accreted mass which is a large fraction of the
total mass of the star and, in turn, a large mass of the circumstellar
disk. To this end, a younger age for the candidate pre-Main Sequence
stars would also reconcile our data with the typical disks observed in
the Galaxy (a few percent of the stellar mass,
e.g. Beckwith~\cite{bec99} and references therein).

In this respect, however, one should also notice that the accretion
rates inferred here are affected by an obvious selection effect, in
that only the strongest excesses, those above the observational
threshold, are detected. In addition, accreting T~Tauri stars are
known to exhibit temporal variation in their spectral features
(e.g. Herbst et al~\cite{her94,her02}). At any given time, then, only
the highest-accreting stars are detected and integrating the measured
value over the lifetime of the star may lead to a gross overestimate
of the total accreted mass.

\subsection{The properties of the young population\label{sec:south_prop}}
Having made the case for the Pre-Main Sequence nature of the \ha\ and
Balmer continuum excess stars, in the next two sections we will derive
global properties of the young population in the South field.

\subsubsection{Spatial distribution of stars\label{sec:south_space}}
Let us now investigate the spatial distribution of the stars in the
field. Since there is no obvious clustering of stars, we have simply
divided the field in four regions, coinciding with the four chips of
the WFPC2.

\begin{figure}[!ht]
\vbox{\psfig{figure=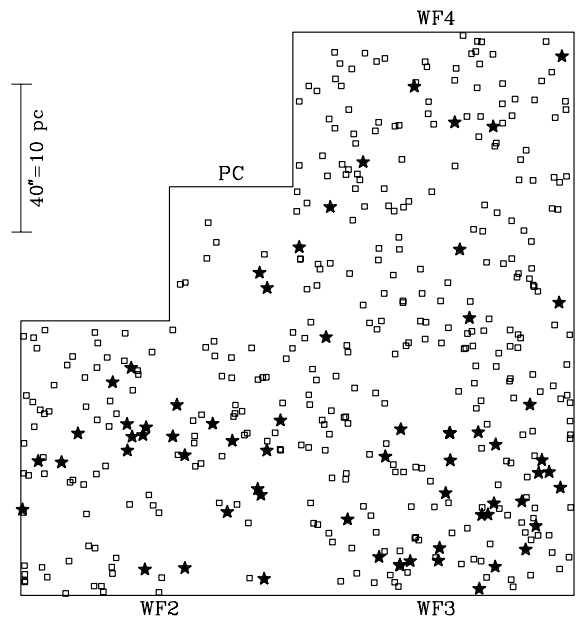,width=\linewidth}}
\caption[]{South Field: spatial distributions of massive ($M\gtrsim
6~M_{\sun}$; star symbol) and Pre-Main Sequence ($M\lesssim
2~M_{\sun}$; open squares) stars belonging to the same young
generation. As in Table~\ref{tab:south_space}, the Pre-Main Sequence
stars included here are only those that can be identified
individually, \ie those with $(m_\mathrm{F675W}-m_\mathrm{F656N})>0.5$.
It appears that the density of hot, massive stars is significantly
higher in WF2 and WF3 than elsewhere.}
\label{fig:south_space}
\end{figure}

As discussed above, in order to study their spatial distribution,
Pre-Main Sequence stars need to be identified on an individual basis,
not just in a statistical sense. Hence, here we will consider only
those objects with $(m_\mathrm{F675W}-m_\mathrm{F656N})>0.5$.  The
spatial distributions of massive and Pre-Main Sequence stars belonging
to the same generation are shown in Figure~\ref{fig:south_space}. The
actual numbers are reported in Table~\ref{tab:south_space} together
with the one of Red Clump stars, which are excellent tracers of the
older field population with ages in excess of a few hundred million years
(e.g. Faulkner \& Cannon~\cite{fau73}).

\begin{table}[!ht]
\begin{minipage}{\linewidth}
\begin{tabular}[b]{*{4}{c}}\hline
Chip & Massive stars   & PMS       & Red Clump    \\ \hline
{\bf PC}  & $2\pm1.4$  & $6\pm2.4$      & $25\pm5$      \\
(PC$\rightarrow$WF$^\mathrm{a}$ & $9.7\pm6.8$ & $28.8\pm11.8$ & $121\pm24$)\\
{\bf WF2} & $24\pm4.9$ & $100\pm10$   & $63\pm7.9$    \\
{\bf WF3} & $29\pm5.4$ & $137\pm11.7$   & $61\pm7.8$    \\
{\bf WF4} & $9\pm3$    & $123\pm11.1$   & $51\pm7.1$    \\ \hline
Total     & $64\pm8$   & $366\pm19.1$ & $200\pm14.1$  \\
(Total$\rightarrow$WF$^\mathrm{b}$ & $19.9\pm2.5$ & $114.4\pm6.0$ & $62.3\pm4.4$)\\
\protect{\footnotetext[1]{Detections in the PC per unit WF area, \ie 4.8
times larger.}}
\protect{\footnotetext[2]{Total detections per unit WF area, \ie 3.2
times smaller than the entire field.}}
\end{tabular}
\end{minipage}
\caption[]{Spatial distribution of different types of stars in the
four chips of the South field. The Pre-Main Sequence stars included
here are only those that can be identified individually, \ie those
with $(m_\mathrm{F675W}-m_\mathrm{F656N})>0.5$. For typographical
convenience we report the Poissonian uncertainty as the square root of
the observed number of stars in each bin.}
\label{tab:south_space}
\end{table}

Most of the \emph{massive stars} are confined to the WF2 and WF3
chips, \ie in the direction of NGC~2050 (see Figure~\ref{fig:dss}),
whereas there is a clear deficiency of them in the PC and WF4
chips. An inspection to Figure~\ref{fig:dss} shows that the
distribution of stars is, indeed, very patchy throughout the entire
region and that the South field happens to be in one of the densest
spots. Interestingly, the distribution of hot stars does not seem to
correlate with the reddening distribution. The number of massive stars
is almost the same in the WF2 and WF3 chips, but the extinction is
appreciably higher in the former than in the latter (see
Figure~\ref{fig:south_ebv} and Table~\ref{tab:south_ebv}), as if the
massive stars in the WF2 chip had not yet dissipated the cloud they
were born from. As a word of caution, however, let us notice that our
method to determine the interstellar extinction does not allow one to
measure the total column density of dust along a line of sight, but,
rather, the one to the star that is used as a beacon. Therefore it is
possible that the true dust distribution is rather uniform across the
two chips, but that stars in the ``more reddened'' one are just more
embedded in the dust cloud. The fact that the dispersions in \ebv\ are
higher where the mean reddening is higher seems to favour this scenario
(see Table~\ref{tab:south_ebv}).

Let us note that the number of low mass Pre-Main Sequence stars we
identify through their strong \ha\ excess is only a lower limit to
their true number. Nevertheless, the spatial distribution of the
detected stars can be taken as representative of the total one. The
distribution of \emph{young low mass stars} belonging to the same
generation as the massive ones is remarkably constant throughout the
field, with the exception of the small area covered by the PC chip. As
a consequence, the ratio of low-mass to massive stars is higher where
the density of massive stars is lower. This is the same trend as
Panagia et al~(\cite{pan00}) found in the field of SN1987A and, as we
will see, it is also the case in the West field. Our results, then,
provide further evidence for the existence of an anti-correlation
between high and low mass young stars, which indicates that star
formation processes for different ranges of stellar masses are rather
different and/or require different initial conditions

The density of \emph{Red Clump stars}, which trace the field
population with ages in excess of a few hundred million years
(e.g. Faulkner \& Cannon~\cite{fau73}), is constant over the three WF
chips and it is only marginally higher in the PC chip. Their density
in the South field is virtually identical to the value measured in the
West one and around SN1987A (Panagia et al~\cite{pan00}). This further
confirms that the older LMC population is uniformly distributed, at
least on the scale of about 100~pc as probed by these three fields.

\subsubsection{The mass function and star formation density
\label{sec:south_mf}}
On the basis of their positions on the HR diagram, we can estimate
fairly accurate masses for the most luminous stars. However, the
individual masses of lower luminosity stars of the young generation
are hard to determine because, taken individually, their displacements
from the MS are comparable to the uncertainties of their effective
temperatures and luminosities.  Therefore, in a discussion of the
Initial Mass Function (IMF), we have to limit ourselves to comparing
the number of stars contained over quite large bins of masses. In
particular, we have 64 stars with $6<M/M_{\sun}<40$ and the 781 stars
with masses between 0.8 and 2~$M_{\sun}$ identified through their \ha\
excess.  Their ratio is consistent with a mass function with a
power-law slope $\alpha=-2.37$, which is indistinguishable from the
``benchmark'' Salpeter~(\cite{salp55}) value of $\alpha=-2.35$.

This result, however, is affected by three sources of incompleteness
that may make the true slope of the stellar mass function considerably
different from the observed one: (1) First, our WFPC2 field is at the
outskirts of a concentration of massive stars (NGC~2050, see
Figure~\ref{fig:dss}) and any mass segregation toward the center of
the cluster could reduce the number of massive stars as compared to
the one of low mass stars, making the observed slope steeper than the
actual one. (2) Second, we can only identify low mass stars showing
strong \ha\ emission that, in Galactic star-forming regions, are only
a fraction of the total number of low mass Pre-Main Sequence
stars. Judging from the results of the West field (see
section~\ref{sec:west}), this is a function of the depth of the \ha\
images and can lead to an underestimate of the true number of Pre-Main
Sequence stars by a factor of two. As a consequence, we can only give
a \emph{lower limit} to the true mass function slope.  (3) Third,
Pre-Main Sequence stars are intrinsically variable (see, for example
Bertout~\cite{bert89}) and their true number is presumably greater,
possibly as much as another factor of 2, than the one determined at
any given time. Again, this makes the observed mass function shallower
than the true one. For example, if the true number of Pre-Main
Sequence stars were $2\times2$ greater than our direct estimate, the
slope of the mass function would become $\alpha=-3$.

The total mass associated to the most recent episode of star formation
can also be estimated, but it will be affected by an even larger
uncertainty because, in addition to a possibly incomplete count of
stars in the range 0.8-2~$M_{\sun}$, we have to extrapolate down to an
unknown lower mass limit. The lowest value of the total mass is
obtained adopting a relatively high value for the lower mass limit
($0.5~M_{\sun}$, Kroupa~\cite{kro02}) and our derived value of the IMF
of $\alpha=-2.37$. Thus we obtain a total mass of about
$3,100~M_{\sun}$. Conversely, using a more canonical lower mass limit
of $0.1~M_{\sun}$ and the high slope estimated above, $\alpha=-3$, the
total mass of the young population would be much larger:
$47,000~M_{\sun}$.  Considering that these are quite extreme values
and that the truth is likely to lie in the middle, we can conclude
that the total mass is likely to be $12,000~M_{\sun}$, to within a
factor of two (1$\sigma$).

Stars of the young generations were formed over a period of time of the
order of 20~Myr.  Therefore, the star formation density in this field
turns out to be  $0.4~M_{\sun}\mathrm{yr}^{-1}\mathrm{kpc}^{-2}$. This
value is intermediate between the typical values found in actively
star-forming spiral disks (up to
$0.1~M_{\sun}\mathrm{yr}^{-1}\mathrm{kpc}^{-2}$) and starburst regions
($1~M_{\sun}\mathrm{yr}^{-1}\mathrm{kpc}^{-2}$ or higher, see, e.g.
Table~1 of Kennicutt~\cite{ken98}). We note that, while  the star
formation activity in the South field is rather high,  still it is a
couple orders of magnitude lower than in 30~Doradus also in the LMC
($100~M_{\sun}\mathrm{yr}^{-1}\mathrm{kpc}^{-2}$ in the central 10~pc,
Kennicutt~\cite{ken98}), the most luminous complex in the Local Group.

\section{The West field\label{sec:west}}
The log of the observations of the West field is reported in
Table~\ref{tab:log_west}. Regrettably, the wide band imaging for this
field is limited to two filters, and, therefore, the discussion will
necessarily be more limited than in the case of the South field.

In the West field we have detected 7,978 stars down to
$m_{\mathrm{F675W}}\simeq24$, 3,416 of which have a mean error in the
two available broadband filters smaller than 0.1 mag:

\begin{equation}
\bar{\delta}_2=\sqrt{\frac{\delta{_\mathrm{F547W}^2}+
\delta{_\mathrm{F675W}^2}}{2}}<0.1
\label{eq:delta2}
\end{equation}

\begin{figure}[!ht]
\vbox{\psfig{figure=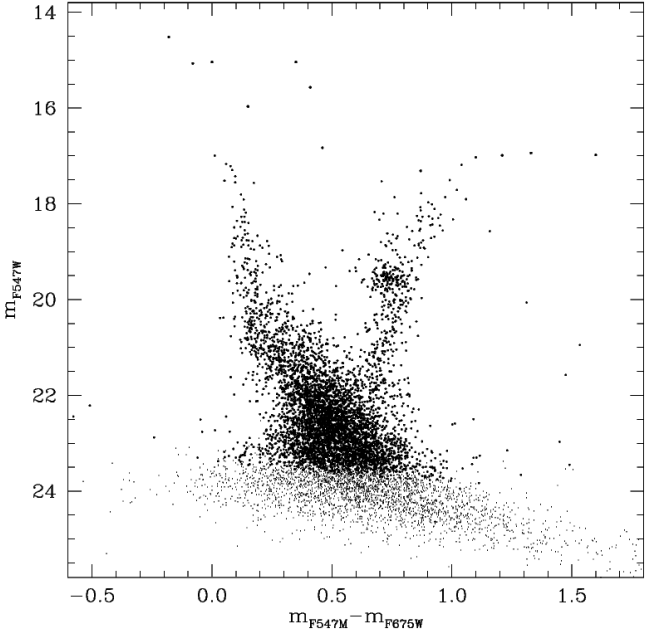,width=\linewidth}}
\caption[]{$m_{\mathrm{F547M}}$~vs~($m_{\mathrm{F547M}}-m_{\mathrm{F675W}}$)
Colour-Magnitude diagram of the 7,978 stars detected in the West field. The
black dots are the 3,416 stars with a mean photometric error in the two bands
($\bar{\delta}_2$, see equation~\ref{eq:delta2}) smaller than 0.1~mag.}
\label{fig:west_cmd}
\end{figure}

This condition, in turn, corresponds to
$m_{\mathrm{F675W}}\lesssim23$. The Colour-Magnitude diagram for the
F547M (V-band like) and F675W (R-band like) filters is shown in
Figure~\ref{fig:west_cmd}.

A visual inspection to the diagram immediately shows the presence of
different generations of stars, covering a wide age range. In
particular, the brightest Main-Sequence stars (about $10~M_{\sun}$)
hint to an age of 20~Myr or lower for the younger population (see
below). As in the case of the South field, the presence of an old
population is revealed by the Red Clump, located at
$m_{\mathrm{F547M}}\simeq19.5$,
$(m_{\mathrm{F547M}}-m_{\mathrm{F675W}})\simeq$0.75, and of a well
developed Red Giant Branch extending up to
$m_{\mathrm{F547M}}\simeq17$ and
$(m_{\mathrm{F547M}}-m_{\mathrm{F675W}})\simeq1$.

\subsection{Reddening and ages\label{sec:west_ebv}}
\begin{figure}[!ht]
\vbox{\psfig{figure=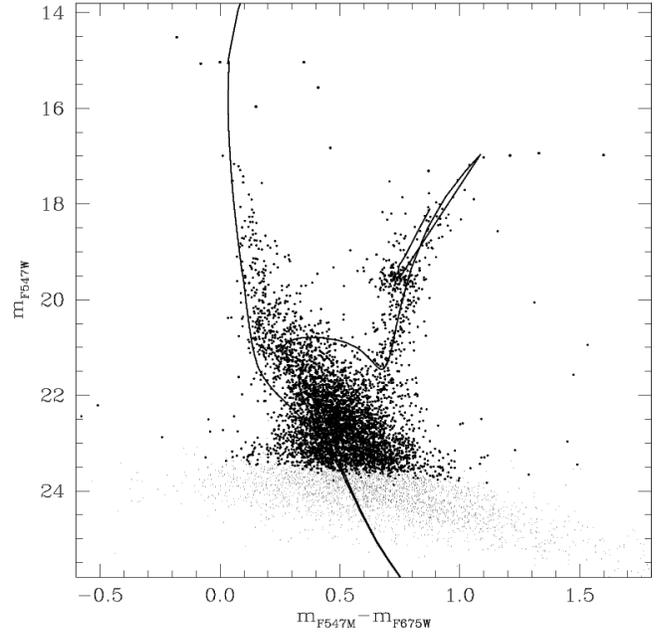,width=\linewidth}}
\caption[]{Same as Figure~{\protect{\ref{fig:west_cmd}}}, but with the
isochrones for 20~Myr and 5~Gyrs (Z=Z$_{\sun}$/3) overplotted to the
data.  The isochrones have been adapted to the WFPC2 bands from the
models of Brocato \& Castellani~(\protect{\cite{bc93}}) and Cassisi et
al~(\protect{\cite{ccs94}}) with the model atmospheres by Bessell et
al~(\protect{\cite{bes98}}). A distance modulus of 18.57 was used
(Romaniello et al~\protect{\cite{rom00}}), yielding an average
reddening of $\ebv=0.25$.}
\label{fig:west_cmd-iso}
\end{figure}

Since the West field was imaged only in two broad bands, we cannot
recover reddening, effective temperature and luminosities for the
individual stars as we did for the South field in
section~\ref{sec:south_ebv}. However, we can still estimate the
\emph{average} reddening by fitting prominent features in the CMD such
as the Main Sequence of the young population and the Red Clump. Using
a distance modulus of $18.57\pm0.05$ (Romaniello et al~\cite{rom00}),
a mean value of $\ebv=0.25$ for the whole field provides a good fit to
the location of both features (see Figure~\ref{fig:west_cmd-iso}).

In addition, Figure~\ref{fig:west_cmd-iso} shows that the Main Sequence
is rather broad,
$\delta(m_{\mathrm{F547M}}-m_{\mathrm{F675W}})\simeq0.2$ at
$m_{\mathrm{F547M}}\simeq20$, reflecting the combined effect of
differential reddening and age spread. The location of the most
luminous, hence youngest, Main Sequence stars indicates an upper limit
of 20~Myr to their age (regrettably, the very bluest star at
$(m_{\mathrm{F547M}}-m_{\mathrm{F675W}})\simeq-0.2$ is so heavily
saturated that its colour is unreliable).

The old population is composed of a mix of stars with ages from
several hundred millions to several billions of years. A
representative isochrone of 5~Gyr is shown in
Figure~\ref{fig:west_cmd-iso}.

\subsection{Pre-Main Sequence stars\label{sec:west_tta}}
\begin{figure}[!ht]
\vbox{\psfig{figure=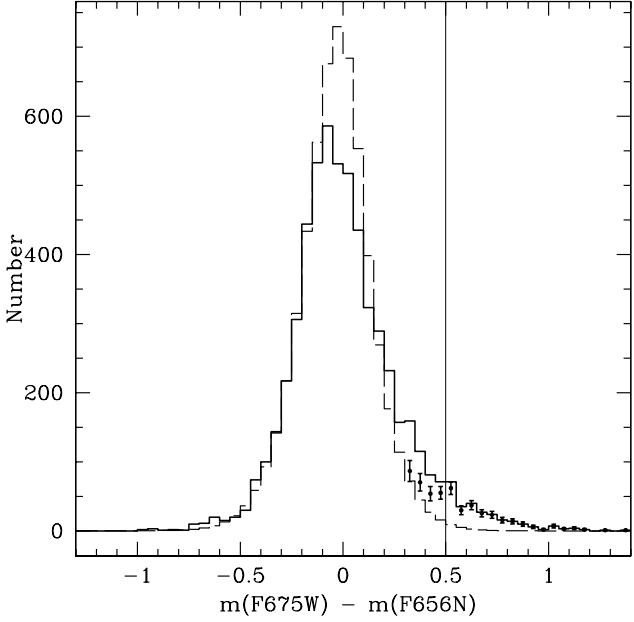,width=\linewidth}}
\caption{As in Figure~\ref{fig:south_ha-ex_all}, but for the stars in
the West field. Also here, the presence of a prominent population of
stars with significant \ha\ excess is evident. The value of the excess
above which stars can be identified individually is marked with a
vertical line.}
\label{fig:west_ha-ex_all}
\end{figure}

Applying the criteria discussed in section~\ref{sec:south_tta} we have
identified $761\pm37$ stars with a statistically significant \ha\
emission (see Figure~\ref{fig:west_ha-ex_all}). Since in the West
field there are only two broad bands available, we could not recover
the effective temperature for the stars. Hence, the expected \ha\
distribution was computed using the relation from the
Kurucz~(\cite{kur93}) models between the only broad-band colour we
have, $(m_{\mathrm{F547M,0}}-m_{\mathrm{F675W,0}})$, and
$(m_{\mathrm{F675W}}-m_{\mathrm{F656N}})$. As it can be seen in
Figure~\ref{fig:west_ha-ex_all}, also in the West field stars with
$(m_{\mathrm{F675W}}-m_{\mathrm{F656N}})>0.5$ can be identified
individually as \ha-emitting objects (264, 35\% of the total of stars
with a statistically significant \ha\ excess).

It is worth noting that, even if the West field represents a sparser
environment than the South field (1.6 times less stars detected in the
former than in the latter), the number of \ha-excess objects detected
is essentially the same in the two fields ($N(\mathrm{South})=781$,
$N(\mathrm{West})=761$).  This is because the \ha\ exposures of the
West field are much deeper than the ones of the South field (a factor
of 2.4 in exposure time, see Tables~\ref{tab:log_south} and
\ref{tab:log_west}). In fact, if we increase the \ha\ photometric
errors in the West field by a factor of $\sqrt{2.4}\simeq1.5$ to make
them comparable to the ones in the South field, our procedure would
identify only 210 candidate Pre-Main Sequence stars. This result
provides an estimate of the uncertainties inherent in the study of an
elusive population such as the one of low-mass Pre-Main Sequence stars
when they are outnumbered by a much older field population.

\begin{figure}[!ht]
\vbox{\psfig{figure=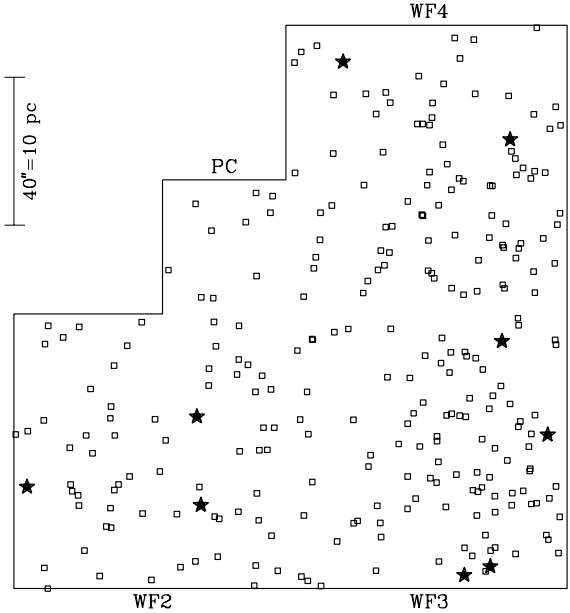,width=\linewidth}}
\caption[]{West field: spatial distributions of massive ($M\gtrsim
6~M_{\sun}$; star symbol) and Pre-Main Sequence ($M\lesssim
2~M_{\sun}$; open squares) stars belonging to the same young
generation.  As in Table~\ref{tab:west_space}, the Pre-Main Sequence
stars included here are only those that can be identified
individually, \ie those with
$(m_\mathrm{F675W}-m_\mathrm{F656N})>0.5$. It is clear that the
candidate Pre-Main Sequence stars are mainly concentrated in the WF3
and WF4 chips, \ie in the direction of SN1987A (see
Figure~\protect{\ref{fig:dss}}).}
\label{fig:west_space}
\end{figure}

The spatial distributions of low mass Pre-Main Sequence and massive
stars in the West field are compared with each other in
Figure~\ref{fig:west_space} and Table~\ref{tab:west_space} (the
Pre-Main Sequence stars are the 264 that can be identified on an
individual basis, \ie $(m_{\mathrm{F675W}}-m_{\mathrm{F656N}})>0.5$).

\begin{table}[!ht]
\begin{minipage}{\linewidth}
\begin{tabular}[b]{*{4}{c}}\hline
Chip & Massive stars  &    PMS   & Red Clump    \\ \hline
{\bf PC}  & $0\pm1$   & $10\pm3.2$   & $8\pm2.8$     \\
(PC$\rightarrow$WF$^\mathrm{a}$ & $0\pm1$ & $48\pm15.4$ & $38.7\pm13.6$)\\
{\bf WF2} & $3\pm1.7$ & $66\pm8.1$  & $49\pm7$      \\
{\bf WF3} & $4\pm2$   & $100\pm10$  & $54\pm7.3$    \\
{\bf WF4} & $2\pm1.4$   & $88\pm9.4$  & $55\pm7.4$    \\ \hline
Total     & $9\pm3$ & $264\pm16.2$  & $166\pm12.9$  \\
(Total$\rightarrow$WF$^\mathrm{b}$ & $2.8\pm0.9$ & $82.5\pm5.1$ & $51.7\pm4.0$)\\
\protect{\footnotetext[1]{Detections in the PC per unit WF area, \ie 4.8
times larger.}}
\protect{\footnotetext[2]{Total detections per unit WF area, \ie 3.2
times smaller than the entire field.}}
\end{tabular}
\end{minipage}
\caption[]{Spatial distribution of different types of stars in the
four chips of the West field.  The Pre-Main Sequence stars included
here are only those that can be identified individually, \ie those
with $(m_\mathrm{F675W}-m_\mathrm{F656N})>0.5$.}
\label{tab:west_space}
\end{table}

Just as in the case of the South field and the region of SN1987A
(Panagia et al~\cite{pan00}), the ratio between low and and high-mass
stars of the same generation exhibits substantial spatial variations
(from 44 in the WF4 chip to roughly half of this value in the WF2
and WF3 chips) mostly due to the variations of the number of PMS among the
various chips.

\subsection{The mass function and star formation density\label{sec:west_mf}}
From Table~\ref{tab:south_space} we see that there are 9 stars with
masses between 6 and $10~M_{\sun}$ and, as discussed in the previous
section, a total of 761 candidate Pre-Main Sequence stars. These numbers
are consistent with a mass function slope of $\alpha\simeq-3.1$.

The same discussion presented in section~\ref{sec:south_mf} can be
repeated for the West field. However, here incompleteness is much less
of a problem, both for massive stars because there is no sign of a
cluster or group of early type stars near this field (see
Figure~\ref{fig:dss}), nor for low-mass PMS stars because for the West
field we have much deeper \ha\ exposures.

Thus, adopting  a slope  $\alpha=-3.1$, and  normalising the mass
function to the number of massive stars present in the field we estimate
a total mass of recently formed stars between $\sim2,300~M_{\sun}$ and
$14,000~M_{\sun}$, depending on whether we assume a lower mass cutoff of
0.5 or $0.1~M_{\sun}$, respectively. 

Taking the geometric mean of the two extremes as a ``most probable"
value of the total mass, $M_{\mathrm{tot}}=5,700~M_{\sun}$, and a
formation time interval comparable to the young star ages, \ie
$\sim20$~Myr, the star formation density in the West field turns out
to be $0.2~M_{\sun}\mathrm{yr}^{-1}\mathrm{kpc}^{-2}$.

\section{Summary and conclusion\label{sec:s&c}}
In this paper we have presented HST-WFPC2 broad and narrow band
imaging of two fields with recent star formation activity in the
Tarantula region of the Large Magellanic Cloud. By means of the
technique pioneered by Romaniello (\cite{rom98}, see also Panagia et
al~\cite{pan00} and Romaniello et al~\cite{rom04}), we have identified
stars stars of 1-2~$M_{\sun}$ with a significant \ha\ emission:
$781\pm33$ in the South field and $761\pm37$ in the West one. The \ha\
emission positively correlates with Balmer continuum excess, indicating
a common physical origin for the two phenomena. The measured \ha\
equivalent widths (up to several tens of \AA; see
Figure~\ref{fig:south_ew}) are far larger than it can be accounted for
by normal chromospheric activity (e.g. Frasca \&
Catalano~\cite{fra94}, White \& Basri~\cite{whi03}).

Given the presence of massive young stars in both fields and the
observed correlation between \ha\ and Balmer continuun excesses , we
interpret the emission objects as Pre-Main Sequence stars. In this
respect, then, they are the equivalent of the Classical TTauri stars
observed in Galactic star-forming regions.

The inherent uncertainty in identifying the Pre-Main Sequence stars
does not allow a precise determination of the mass function of the
young population. We have discussed the different effects that bias
its determination and placed a firm lower limit to its slope based
solely on \ha-excess stars of $\alpha=-2.37$ and $-3.1$ in the South
and West field, respectively. The real mass function could be steeper
should the selection criterion we have adopted miss a significant
number of stars.

Also the determination of the star formation rate associated with the
young generation of stars is affected by the selection criteria
applied to identify the candidate Pre-Main Sequence stars, with
variations of a factor of a few. Our best estimates for the star
formation density are of
$\sim0.4~M_{\sun}\mathrm{yr}^{-1}\mathrm{kpc}^{-2}$ in the South field
and $\sim0.2~M_{\sun}\mathrm{yr}^{-1}\mathrm{kpc}^{-2}$ in the West
field.  These values are intermediate between what is found for
actively star-forming spiral disks (less than
$0.1~M_{\sun}\mathrm{yr}^{-1}\mathrm{kpc}^{-2}$) and starbursts
($1~M_{\sun}\mathrm{yr}^{-1}\mathrm{kpc}^{-2}$ or more;
Kennicutt~\cite{ken98}).

The relative spatial distribution of equally young stars with
different masses, however, is not affected by the bias on the
selection criteria for Pre-Main Sequence stars and provides clues to
the mechanisms that lead to star formation. For both fields the
spatial location of the low mass Pre-Main Sequence stars does not
follow the one of the massive stars of the same young age. This
further confirms the results found by Panagia et al~(\cite{pan00}) for
the field of SN1987A, also is in the Tarantula region (see
Figure~\ref{fig:dss}). \emph{The almost anti-correlation of spatial
distributions of high and low mass stars of a coeval generation
indicates that star formation processes for different ranges of
stellar masses are rather different and/or require different initial
conditions}. An important corollary of this result is that the very
concept of an Initial Mass Function seems not to have validity in
detail, but may rather be the result of a random process, so that it
could make sense to talk about an average IMF over a suitably large
area, in which all different star formation processes are concurrently
operating.

\appendix
\section{The relation between EW(H{\boldmath ${\alpha}$}) and {\boldmath
$m(\mathrm{F675W})-m(\mathrm{F656N})$}\label{sec:app_ew-dm}}

In order to compute the relation between the colour excess
$m(\mathrm{F675W})-m(\mathrm{F656N})$ and the equivalent width of the
\ha\ line, let us begin by defining $T_R(\lambda)$ as the total response of the
system in the broad-band filter R (telescope+filter+detector) and
$f(\lambda)=f_c+f_l(\lambda)$ the total flux from the source
(continuum+line emission). If the continnum is assumed to be flat
across the broad band, the \emph{detected} flux in it is:

\begin{eqnarray*}
f_{R}&=&\int_{R} T_{R}(\lambda) f(\lambda) d\lambda=\\
&=&f_c\int_{R}T_{R}(\lambda)
d\lambda+T_{R}(\ha)\int_{R} f_l(\lambda) d\lambda
\end{eqnarray*}
where we have further assumed the line to be much narrower than the
filter response curve across it so that, in the last term, this latter
quantity can be replaced with its value at the central wavelength of
the line, $T_{R}(\ha)$. From the definition of equivalent width (EW),
it follows that:

\begin{displaymath}
\int_{R} f_l(\lambda) d\lambda=\mathrm{EW(\ha)}\cdot f_c
\end{displaymath}
and, hence:

\begin{equation}
f_{R}=f_c\int_{R}T_{R}(\lambda)d\lambda+T_{R}(\ha)\cdot f_c\cdot
\mathrm{EW(\ha)}
\label{eq:f_BB}
\end{equation}
A similar equation holds for the measured flux in the \ha\ narrow-band
filter:

\begin{equation}
f_{\ha}=f_c\int_{\ha}T_{\ha}(\lambda)d\lambda+T_{\ha}(\ha)\cdot f_c\cdot
\mathrm{EW(\ha)}
\label{eq:f_NB}
\end{equation}
where $T_{\ha}(\lambda)$ is the response curve of the narrow band filter.

Finally, transforming fluxes to magnitudes leads to the relation we are
looking for, although in an implicit form:

\begin{equation}
m_\mathrm{R}-m_{\ha}=-2.5\log(f_{R}/f_{\ha})+\Delta(ZP)
\label{eq:dmag}
\end{equation}
where $\Delta(ZP)$ is the difference between the \emph{instrumental}
zero points in the broad and narrow band. In our case, we have used the
WFPC2 filters F675W as R and F656N as \ha\ and $\Delta(ZP)=4.312$ (Whitmore
\cite{whit95}).

\begin{figure}[!ht]
\vbox{\psfig{figure=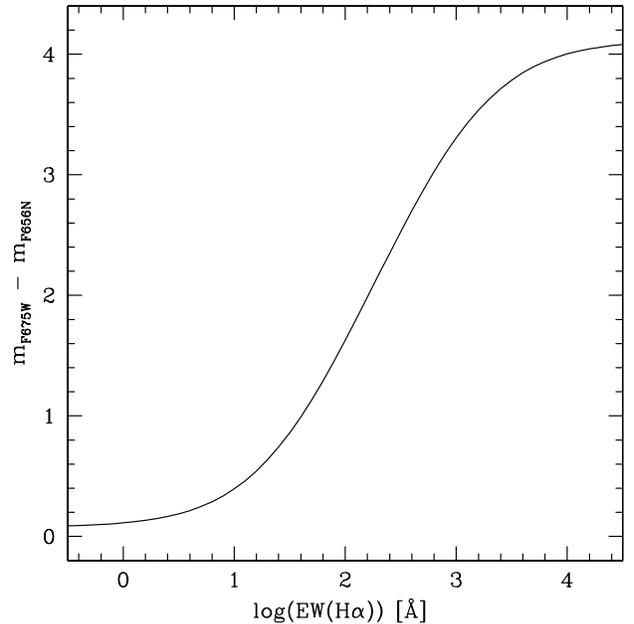,width=\linewidth}}
\caption[]{Colour excess $m_\mathrm{F675W}-m_\mathrm{F656N}$ as a function
of the \ha\ equivalent width computed from equations~(\ref{eq:f_BB}),
(\ref{eq:f_NB}) and (\ref{eq:dmag}).}
\label{fig:dmag_ew}
\end{figure}

The magnitude difference $(m_\mathrm{F675W}-m_\mathrm{F656N})$ as a
function of the \ha\ equivalent width as computed by combining
equations~(\ref{eq:f_BB}), (\ref{eq:f_NB}) and (\ref{eq:dmag}) is
shown in Figure~\ref{fig:dmag_ew}.

\begin{acknowledgements}
We warmly thank the anonymous referee of an early version of the paper
for many comments that lead to considerably improving the presentation
of our results, and the one of the final draft for many useful
suggestions.
\end{acknowledgements}

\listofobjects


\begin{thebibliography}{}
\bibitem[1972]{ard72} Ardeberg A., Brunet J.P., Maurice E., \& Prevot L.
  1972, \aaps, 6, 249.
\bibitem[1999]{bec99} Beckwith, S.V.W. 1999, in The Origin of Stars
  and Planetary Systems, eds. C.J. Lada and N.D. Kylafis (Dordrecht:
  Kluwer Academic Publishers), 579
\bibitem[1989]{bert89} Bertout, C. 1989, \araa, 27, 351.
\bibitem[1991]{bes91} Bessell, M.S. 1991, \aap, 242, L17.
\bibitem[1998]{bes98} Bessell, M.S., Castelli, F., \& Plez, B. 1998, \aap,
  333, 231 (erratum 337, 321).
\bibitem[1993]{bc93} Brocato, E., \& Castellani, V. 1993, \apj, 410, 99.
\bibitem[2002]{cal02} Calvet, N., Hartmann, L., \& Strom, S.E. 2002,
  in Protostars and Planets IV, ed. V. Mannings, A.P. Boss, \&
  S.S. Russell (Tucson: Univ. Arizona Press), 377.
\bibitem[2005]{cal05} Calvet, N., Brice\~{n}o, C., Hern\'andez, J. et al,
  \aj, 129, 935.
\bibitem[1993]{ccs94} Cassisi, S., Castellani, V., \& Straniero, O. 1994,
  \aap, 282, 753.
\bibitem[1995]{coo95} Cool, A.M, \& King, I.R. 1995, in Calibrating HST: Post
 Servicing Mission, eds. A. Koratkar and C. Leitherer (Baltimore: STScI), 290.
\bibitem[1980]{con80} Conover, W.J. 1980, ``Practical Nonparametric
  Statistics'', $2^{nd}$ edition, (New York:Wiley).
\bibitem[2002]{wit02}  de Wit, W.J., Beaulieu, J.P., \& Lamers, H.J.L.M.
  2002, \aap, 395, 829.
\bibitem[2005]{wit05} de Wit, W.J., Beaulieu, J.P., Lamers, H.J.L.M.,
  Coutures, C., \& Meeus, G. 2005, \aap, 432, 619.
\bibitem[1973]{fau73} Faulkner, D.J., \& Cannon, R.D. 1973, \apj, 180, 435.
\bibitem[1995]{fer95} Fern\'andez, M., Ortiz, E., Eiroa, C., \&
  Miranda, L.F. 1995, \aaps, 114, 439.
\bibitem[1988]{fit88} Fitzpatrick, E.L. 1988, \apj, 335, 703.
\bibitem[1994]{fra94} Frasca, A., \& Catalano, S. 1994, \aap, 284, 883.
\bibitem[1994]{gil94} Gilliland, R.L. 1994, \apj, 435, L63.
\bibitem[1990]{gil90} Gilmozzi, R. 1990, Core Aperture Photometry with the
  WFPC, STScI Instrum. Rep. WFPC-90-96 (Baltimore: STScI).
\bibitem[1990]{gilm94} Gilmozzi, R., Kinney, E.K.,
   Ewald, S.P., Panagia, N., Romaniello, M. 1994,  \apj, 435, L43.
\bibitem[1998]{gul98} Gullbring, E., Hartmann, L., Briceno, C., \& Calvet, N.
  1998, \apj, 492, 323.
\bibitem[2004]{wfpc2} Heyer, I., Biretta, J. et al. 2004, WFPC2 Instrument
  Handbook, Version 9.0 (Baltimore: STScI).
\bibitem[1998]{ken98} Kennicutt, R.C. 1998, \araa, 36,189.
\bibitem[2002]{kro02} Kroupa, P. 2002, Science, 295, 82.
\bibitem[1993]{kur93} Kurucz, R.L. 1993, ATLAS9 Stellar Model Atmosperes
  Programs and $2~\mathrm{km s}^{-1}$ grid (Kurucz CD-ROM No.13).
\bibitem[1999]{lam99} Lamers, H.J.G.L.M., Beaulieu, J. P., \& de Wit, W. J.
  1999, \aap, 341, 827.
\bibitem[1994]{her94} Herbst, W., Herbst, D.K., \& Grossman, E.J. 1994
  \aj, 108, 1906.
\bibitem[2002]{her02} Herbst, W, Bailer-Jones, C.A.L., Mundt, R.,
  Meisenheimer, K., \& Wackermann, R. 2002, \aap, 396, 513.
\bibitem[1988]{hod88} Hodge, P.W. 1988, \pasp, 100, 1051.
\bibitem[1996]{mad96} Madau, P., Ferguson, H.C., Dickinson, M.E., et al 1996,
  \mnras, 283, 1388.
\bibitem[1995]{mas95} Massey, P., Johnson, K.E., \& Degioia-Eastwood, K.
  1995, \apj, 454, 151.
\bibitem[1988]{mat88} Mateo, M. 1988, \apj, 331, 261.
\bibitem[1996]{padg96} Padgett, D.L. 1996, \aj, 471, 874.
\bibitem[1991]{ps91} Palla, F., \& Stahler, S. 1991, \apj, 360, 47.
\bibitem[2000]{pan00} Panagia, N., Romaniello, M., Scuderi, S., \& Kirshner,
  R.P. 2000, \apj, 539, 197 (Paper I).
\bibitem[1999]{pei99} Pei, Y.C., Fall, S.M., \& Hauser, M.G. 1999, \apj, 522,
  604.
\bibitem[2004]{rob04} Robberto, M., Song, J., Mora Carrillo, G., Beckwith,
  S.V.W., Makidon, R.B., \& Panagia, N. \apj, 606, 952.
\bibitem[1998]{rom98} Romaniello, M. 1998, Ph.D. Thesis, Scuola Normale
  Superiore, Pisa.
\bibitem[2000]{rom00} Romaniello, M., Salaris, M., Cassisi, S., \& Panagia, N.
  2000, \apj, 530, 738.
\bibitem[2002]{rom02} Romaniello, M., Panagia, N., Scuderi, S., \& Kirshner,
  R.P. 2002, \aj, 123, 91.
\bibitem[2004]{rom04} Romaniello, M., Robberto, M., \& Panagia, N. 2004, \apj,
  608, 220.
\bibitem[1955]{salp55} Salpeter, E.E. 1955, \apj, 121, 161.
\bibitem[1969]{sand69} Sanduleak, N. 1969, Contr. Cerro Tololo Interam. Obs,
  No. 89.
\bibitem[1993]{sch93} Schaerer, D., Meynet, G., Maeder, A., \& Schaller, G.
  1993, \aaps, 98, 523.
\bibitem[1982]{sk82} Schmidt-Kaler, T. 1982, in Landolt-Bornstein: Numerical
  Data and Functional Relationships in Science and Technology, vloume 2b
  (Berlin:Springer-Verlag), 297.
\bibitem[1991]{sch91} Schwering, P.B.W., \& Israel, F.P. 1991, \aap, 246, 231.
\bibitem[1996]{scud96} Scuderi, S., Panagia, N., Gilmozzi, R., Challis, P.M.,
  Kirshner, R.P., 1996, \apj, 465, 956. 
\bibitem[1997]{siess97} Siess L., Forestini M., \& Dougados, C. 1997, \aap,
  324, 556.
\bibitem[2000]{sir00} Sirianni, M., Nota, A., Leitherer, C., De Marchi, G.,
  \& Clampin, M. 2000, \apj, 533, 203.
\bibitem[2001]{mar01} van der Marel, R.P., \& Cioni, M.-R.L. 2001,
  \aj, 122, 1807.
\bibitem[2001]{wic01} Wichmann, R., Schmitt, J.H.M.M. \& Krautter, J.
  2001, \aap, 380, L9.
\bibitem[2003]{whi03} White, R.J, \& Basri, G. 2003, \apj, 582, 1109.
\bibitem[1995]{whit95} Whitmore, B. 1995, in Calibrating Hubble Space
  Telescope: Post Servicing Mission, eds. A. Koratkar and C. Leitherer
  (Baltimore: STScI), 269.
\bibitem[1999]{zar99} Zaritsky, D. 1999, \aj, 118, 2824.
\end{thebibliography}
\end{document}